\documentclass[%
 reprint,
 amsmath,amssymb,
 aps,
]{revtex4-2}

\usepackage{graphicx}
\usepackage{dcolumn}
\usepackage{bm}

\begin{document}

\preprint{APS/123-QED}

\title{Mechanical response of packings of non-spherical particles: A case study of 2D packings of circulo-lines}

\author{Jerry Zhang,\textit{$^{1}$} Kyle VanderWerf,\textit{$^{2,5}$} Chengling Li,\textit{$^{1,6}$} Shiyun Zhang,\textit{$^{1,7}$} Mark D. Shattuck,\textit{$^{3}$} and Corey S. O'Hern\textit{$^{1,2,4}$}}
\affiliation{\textit{$^{1}$~Department of Mechanical Engineering and Materials Science, Yale University, New Haven, Connecticut 06520, USA.}}
\affiliation{\textit{$^{2}$~Department of Physics, Yale University, New Haven, Connecticut, 06520, USA.}}
\affiliation{\textit{$^{3}$~Benjamin Levich Institute and Physics Department, The City College of New York, New York, New York 10031, USA.}}
\affiliation{\textit{$^{4}$~Department of Applied Physics, Yale University, New Haven, Connecticut 06520, USA.}}
\affiliation{\textit{$^{5}$~MIT Lincoln Laboratory, Lexington, MA 02421, USA.}}
\affiliation{\textit{$^{6}$~Department of Physics, Emory University, Atlanta, GA 30322, USA.}}
\affiliation{\textit{$^{7}$~Department of Physics, University of Science and Technology of China, Hefei, Anhui 230026, China.}}

\date{\today}

\begin{abstract}
We investigate the mechanical response of jammed packings of circulo-lines, interacting via purely repulsive, linear spring forces, as a function of pressure $P$ during athermal, quasistatic isotropic compression. Prior work has shown that the ensemble-averaged shear modulus for jammed disk packings scales as a power-law, $\langle G(P) \rangle \sim P^{\beta}$, with $\beta \sim 0.5$, over a wide range of pressure.  For packings of circulo-lines, we also find robust power-law scaling of $\langle G(P)\rangle$ over the same range of pressure for aspect ratios ${\cal R} \gtrsim 1.2$.  However, the power-law scaling exponent $\beta \sim 0.8$-$0.9$ is much larger than that for jammed disk packings. To understand the origin of this behavior, we decompose $\langle G\rangle$ into separate contributions from geometrical families, $G_f$, and from changes in the interparticle contact network, $G_r$, such that $\langle G \rangle = \langle G_f\rangle + \langle G_r \rangle$.  We show that the shear modulus for low-pressure geometrical families for jammed packings of circulo-lines can both increase {\it and} decrease with pressure, whereas the shear modulus for low-pressure geometrical families for jammed disk packings only decreases with pressure.  For this reason, the geometrical family contribution $\langle G_f \rangle$ is much larger for jammed packings of circulo-lines than for jammed disk packings at finite pressure, causing the increase in the power-law scaling exponent. 
\end{abstract}

\maketitle


\section{\label{sec:level1}Introduction}

Granular materials represent fascinating examples of nonequilibrium physical systems that display complex, collective behavior~\cite{rmp,phys_jamming}, such as stick-slip motion~\cite{cain}, shear banding~\cite{fenistein}, and segregation~\cite{segregation}.  They are composed of discrete, macroscopic grains that interact via dissipative, frictional contact interactions.  Granular materials occur in many important contexts, including numerous geological processes~\cite{benzion,benzion2}, food and consumer product processing~\cite{chen,muzzio}, and robotics applications~\cite{brown,kramer}.  Because granular media are highly dissipative, the grains do not move unless they are driven by gravity, fluid flow, or the system boundaries.  When granular systems are compressed to sufficiently large packing fractions, they become jammed and possess a nonzero static shear modulus and other soid-like properties~\cite{epitome}.

While most dry granular materials are composed of frictional, non-spherical grains, numerous computational and theoretical studies of dry granular packings have focused on the soft-particle model in which frictionless, spherical particles interact via the pairwise, purely repulsive potential energy~\cite{dd}: $U(r_{ij}) \propto (1-r_{ij}/\sigma_{ij})^{\alpha} \Theta(1-r_{ij}/\sigma_{ij})$, where $r_{ij}$ is the separation between the centers of mass of particles $i$ and $j$, $\sigma_{ij}$ is their average diameter, the exponents $\alpha=2$ and $5/2$ are set for purely repulsive linear and Hertzian spring interactions~\cite{gm}, and $\Theta(.)$ is the Heaviside step function that ensures the potential energy is nonzero only when the particles are in contact. For frictionless, spherical particles, the jamming transition occurs when the number of interparticle contacts $N_c$ equals or exceeds the isostatic value $N_c^0=dN - d+1$, where $d$ is the spatial dimension and $N$ is the number of non-rattler particles~\cite{witten}. When the system is compressed above jamming onset, the pressure increases from zero and the jammed packing develops nonzero bulk and shear moduli. 

A hallmark of the jamming transition in static packings of frictionless, spherical particles is that the ensemble-averaged contact number and shear modulus $\langle G \rangle$ scale as a power-law in the pressure $P$~\cite{epitome,makse}, above a characteristic pressure $P^{**}$ that decreases with increasing system size, as the packings are isotropically compressed above jamming onset~\cite{goodrich}.  For example, $\langle G \rangle \sim P^{1/2}$ for $P > P^{**}$ for packings of spherical particles with purely repulsive, linear spring interactions. Several previous experimental studies of compressed emulsions~\cite{mason,brujic} and packings of thin granular cylinders~\cite{behringer} have also found power-law scaling of the contact number and shear modulus with pressure during isotropic compression. In previous computational studies of packings of frictionless, spherical particles~\cite{Kyle_PRL}, we showed that there are two important contributions to the ensemble-averaged shear modulus: $G_f$ from geometrical families and $G_r$ from changes in the interparticle contact network during compression.  For isotropic compression, jammed packings within a geometrical family are mechanically stable packings with different pressures that are related to each other by continuous, quasistatic changes in packing fraction with no changes in the interparticle contact network. $G_r$ includes discontinuities in the shear modulus that arise from point and jump changes in the contact network~\cite{PJ_jump_changes,plasticity}. Point changes involve the addition or removal of a single interparticle contact (or multiple contacts when a rattler particle is added or removed from the contact network) without significant particle motion. Jump changes are caused by mechanical instabilities and typically involve multiple changes in the contact network and collective particle motion.

For frictionless, spherical particles, the shear modulus along a geometrical family {\it decreases} with increasing pressure during isotropic compression (and no shearing of the boundaries). For systems with purely repulsive linear spring interactions, $G_f$ decreases roughly linearly with pressure, $G_f(P)/G_0 \sim 1 - P/P_0$, where $G_0$ is the shear modulus at $P=0$ and $P_0$ is the pressure at which $G_f(P)=0$, although there are deviations from this simple form as the system approaches contact network changes~\cite{Philip_Herzian,Kyle_PRL}. Thus, the increase in the ensemble-averaged shear modulus, $\langle G\rangle$, with pressure for packings of spherical particles arises from $G_r$, where changes in the contact network cause discontinuous upward jumps in the shear modulus. We will show below that the sign of the second, pressure-dependent term in $G_f$ (i.e. whether $G_f$ increases or decreases with pressure) is determined by the curvature of geometrical families in the strain direction. 

Several prior computational studies have shown that the form of $\langle G(P) \rangle$ for jammed packings depends on the particle shape~\cite{mailman,Shreck_dimer_ellipse}. For example, the pressure-depenent shear modulus for packings of frictionless ellipse-shaped particles with repulsive linear spring interactions scales as $\langle G(P) \rangle \sim P^{\beta}$, where $\beta \sim 1$ over a wide range of pressure, $P^{**} < P < P^*$, $P^{**} \sim N^{-2}$, and $P^*$ does not depend strongly on system size. For $P>P^*$, the power-law scaling crosses over to $\langle G \rangle \sim P^{1/2}$, as found for jammed packings of spherical particles~\cite{goodrich}.  However, the ensemble-averaged shear modulus for jammed packings of dimer-shaped particles (and other composite particles formed from bonded spherical particles) scales as $\langle G \rangle \sim P^{1/2}$ over the same range of pressures and for the same aspect ratios as those studied for packings of smooth, ellipse-shaped particles~\cite{Shreck_dimer_ellipse}. Since packings of frictionless ellipse-shaped particles are hypostatic at jamming onset, while packings of dimer-shaped particles are isostatic at jamming onset, it is possible that the change in the power-law scaling exponent from $\beta =0.5$ to $\sim 1$ is related to the presence of low-frequency quartic modes in the vibrational response for packings of ellipse-shaped and other non-spherical particles~\cite{Kyle_nonsphere}. 

In this work, we address the question of what determines the power-law scaling exponent $\beta$ for packings of non-spherical particles.  Does $G_f$ decrease with pressure for packings of non-spherical particles? Are the frequency of contact network changes and the magnitude of the discontinuities in the shear modulus at contact changes different from those for packings of spherical particles?  As a case study, we investigate the pressure-dependent mechanical response of jammed packings of circulo-lines interacting via purely repulsive, linear spring interactions in two spatial dimensions (2D) as a function of aspect ratio.  

We find several key results for the mechanical response of jammed packings of circulo-lines. First, the curvature of the variation of packing fraction with strain for geometrical families can be either {\it negative or positive}, and thus the shear modulus of geometrical families can either {\it increase or decrease} with pressure: $G_f/G_0 \sim 1 \pm P/P_0$ to linear order in pressure. We derive an exact expression for the pressure-dependent shear modulus of jammed packings and show that near jamming onset it can be approximated as $G_f \sim -\frac{1}{\phi}\left(\frac{dP}{d\gamma}\right)_\phi\left(\frac{d\phi}{d\gamma} \right)_P - \frac{P}{\phi}\left(\frac{\left(\frac{d\phi}{d\gamma}\right)_P}{d\gamma} \right)_\phi$~\cite{Sheng_stress_anisotropy}, where $\gamma$ is the shear strain. The first term tends to a constant, $G_0 >0$, in the zero-pressure limit and the sign of the coefficient of the second term (that is roughly linear in $P$) is determined by the curvature of geometric families in the $\phi$-$\gamma$ plane. Second, we decompose $G_f = G_a - G_{na}$ for each first, low-pressure geometrical family into its affine and non-affine contributions. $G_a$ gives the response of the system to a globally affine change of the particle positions and system boundaries, while $G_{na}$ also includes particle motion in response to potential energy minimization.  We find that the non-affine term plays an important role in determining $G_f$. In particular, the non-affine contribution can cause $G_f$ to increase with pressure, which does not occur in jammed packings of spherical particles.
We also calculate the ensemble-averaged shear modulus $\langle G\rangle$ versus $P$ for jammed packings of circulo-lines over a range of aspect ratios ${\cal R}$ and system sizes. For packings of circulo-lines, we find that $\langle G \rangle \sim P^{\beta}$, where $\beta \sim 0.8$-$0.9$, over a range of pressures $P^{**} < P < P^*$, where $P^{**} \sim N^{-2}$ and $P^*$ decreases as ${\cal R} \rightarrow 1$ and does not depend strongly on system size.  We find that the finite fraction of geometrical families with negative curvature in the $\phi$-$\gamma$ plane causes the power-law exponent $\beta$ for $\langle G\rangle$ versus $P$ to increase compared to that for jammed packings of spherical particles.  

The remainder of the article is organized as follows. In Sec.~\ref{methods}, we describe the interparticle potential and types of contacts that occur between pairs of circulo-lines, the numerical methods used to generate jammed packings of circulo-lines, and formulas for calculating the pressure, shear stress, and shear modulus. In Sec.~\ref{results}, we describe the results concerning geometrical families, non-affine contributions to the shear modulus, point and jump changes, and the ensemble-averaged shear modulus for jammed packings of circulo-lines. Finally, in Sec.~\ref{conclusions}, we provide the conclusions and point to promising future research directions.  We also include an appendix that gives explicit expressions for the affine shear modulus for packings of circulo-lines with purely repulsive, linear spring interactions.

\section{Methods}
\label{methods}

\begin{figure}[h!]
\centering
\includegraphics[height=19cm]{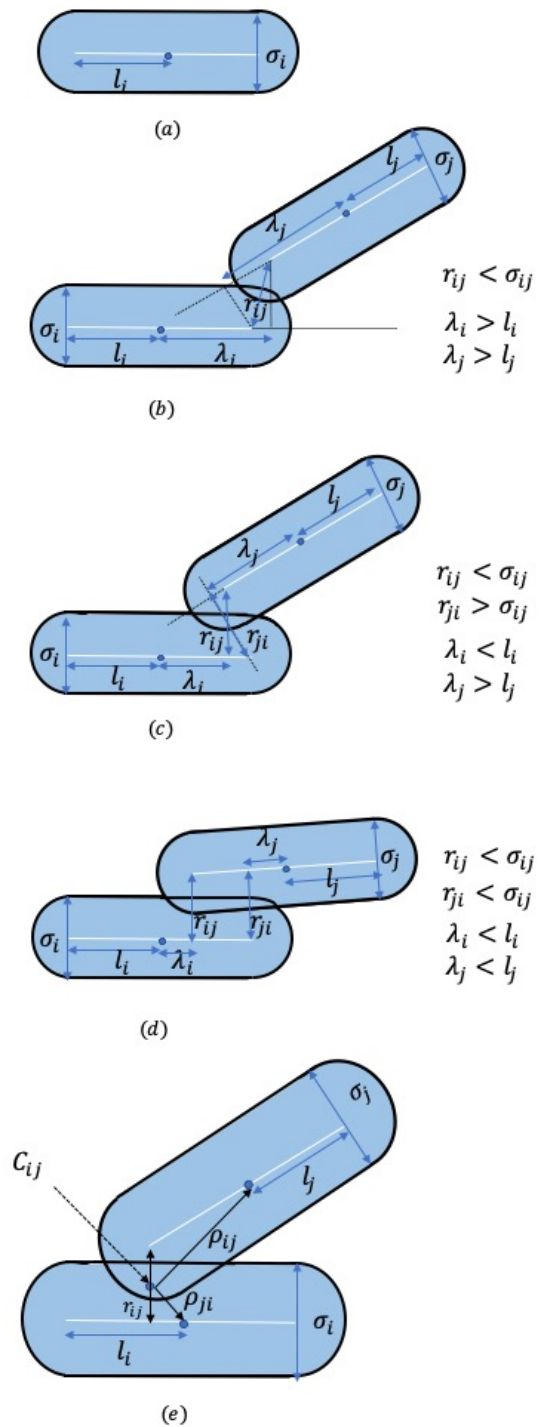}
\caption {(a) A single circulo-line with middle segment length $2l_i$ and end cap diameter $\sigma_i$. The three types of contacts between a pair of circulo-lines $i$ and $j$: (b) an end-end contact, (c) an end-middle contact, and (d) a middle-middle contact. Note that $\sigma_{ij} = (\sigma_i+\sigma_j)/2$ and the distances $r_{ij}$, $r_{ji}$, $\lambda_i$, and $\lambda_j$ are defined in the main text. (e) The effective point of contact $C_{ij}$ between two overlapping circulo-lines $i$ and $j$ is located at the midpoint of $r_{ij}$. The vector ${\vec \rho}_{ij}$ points from $C_{ij}$ to the center of circulo-line $j$.}
\label{fgr:Shape}
\end{figure}

A circulo-line is the set of points that are equally distant from a line segment; it is thus a 2D shape composed of a rectangular middle region capped by two semi-circles on both ends.  Fig.~\ref{fgr:Shape} (a) shows a circulo-line (labeled $i$) with diameter of the semi-circles, $\sigma_i$, and length of the middle line segment, $2 l_i$. We will refer to the end points of the middle line segment as the foci. The aspect ratio of a circulo-line is $\mathcal{R}=(\sigma_i+2 l_i)/\sigma_i$, and ${\cal R} =1$ in the limit that the particle becomes coincident semi-circles.  The asphericity of a circulo-line is $\mathcal{A}=p_i^2/4\pi a_i=\frac{(2\pi+4(\mathcal{R}_i-1))^2}{4\pi(\pi+4(\mathcal{R}_i-1))}$, where $p_i$ is the perimeter and $a_i$ is the area of the circulo-line.  In these studies, we vary ${\cal R}-1$ and ${\cal A}-1$ over the ranges $10^{-2}$ to $2$ and $10^{-5}$ to $0.5$, respectively. We study bidisperse packings of circulo-lines to inhibit positional and orientational order.  We consider packings in rhombic boxes with edge length $L$, periodic boundary conditions in the $x$- and $y$-directions, and $N/2$ large and $N/2$ small particles with end cap diameter ratio $\sigma_l/\sigma_s=1.4$, but the same mass $m$ and aspect ratio ${\cal R}$. To investigate the effect of system size, we studied packings with $N=64$, $128$, $256$, and $512$.

In Fig.~\ref{fgr:Shape} (b)-(d), we show that there are three ways in which two circulo-lines can make contact (or make small overlaps) with each other: end-end, end-middle, and middle-middle contacts. Fig.~\ref{fgr:Shape} (b) shows an end-end contact between circulo-lines $i$ and $j$. In this case, the relevant separation $r_{ij}$ between the circulo-lines is the separation between the two closest foci on $i$ and $j$.  Another important distance is the separation $\lambda_i$ between the center of circulo-line $i$ and the point at which the line from the closest foci on $j$ that is perpendicular to the axis of circulo-line $i$ intersects the axis of $i$. For an end-end contact, the following three conditions must be satisfied: $r_{ij} < \sigma_{ij} = (\sigma_i+\sigma_j)/2$, $\lambda_i > l_i$, and $\lambda_j > l_j$. 

Fig.~\ref{fgr:Shape} (c) shows an end-middle contact where the end of circulo-line $j$ is in contact with the middle region of circulo-line $i$. For an end-middle contact, the relevant separation is the distance $r_{ji}$ between the closest focus on $i$ and the axis of $j$. Similarly, we can define the separation $r_{ij}$, but $r_{ji} \ne r_{ij}$.  For an end-middle contact, the following four conditions must be satisfied: $r_{ij}< \sigma_{ij}$, $r_{ji} > \sigma_{ij}$, $\lambda_i <l_i$, and $\lambda_{j} > l_j$.

Fig.~\ref{fgr:Shape} (d) shows two circulo-lines for which the two middle regions are in contact. We model middle-middle contacts as two end-middle contacts, so that when a middle-middle contact becomes an end-end contact (from Fig.~\ref{fgr:Shape} (d) to (b)) or end-middle contact (from Fig.~\ref{fgr:Shape} (d) to (c)) due to small changes in the positions and orientations of the circulo-lines, the interparticle potential energy and forces are continuous. Thus, for a middle-middle contact, the following four conditions must be satisfied:  $r_{ij}< \sigma_{ij}$, $r_{ji} < \sigma_{ij}$, $\lambda_i <l_i$, and $\lambda_{j} < l_j$.

For all three types of contacts, we use the pairwise, purely repulsive, linear spring potential 
\begin{equation}
\label{energy}
    U(r_{ij}) = \frac{\epsilon}{2} \left(1-\frac{r_{ij}}{\sigma_{ij}} \right)^{2} \Theta \left( 1-\frac{r_{ij}}{\sigma_{ij}} \right),
\end{equation}
where $\epsilon$ is the characteristic energy scale of the interaction and $\Theta(.)$ is the Heaviside step function, ensuring that pairs of circulo-lines do not interact when then are not in contact. The total potential energy, $U=\sum_{i>j} U(r_{ij})$, and pair force, ${\vec f}_{ij} = (dU/dr_{ij}) {\hat r}_{ij}$, are continuous as a function of the coordinates ${\vec r}_i$ of the centers of mass and orientations $\theta_i$ of the circulo-lines.  We use $\epsilon$, $\epsilon/\sigma_s$, and $\epsilon/\sigma^2_s$ for the units of energy, force, and stress. 

We employ the Love expression~\cite{Love_expression} to calculate the stress tensor: 
\begin{equation}
\label{lovesigma}
    \Sigma_{\alpha\beta} = \frac{1}{2L^2} \sum^N_{i,j=1} (f_{ij\alpha}\rho_{ij\beta}+f_{ij\beta}\rho_{ij\alpha}),
\end{equation}
where $f_{ij\alpha}$ is the $\alpha$-component of the force on particle $i$ due to particle $j$ and $\rho_{ij\beta}$ is the $\beta$-component of the vector pointing from the effective contact point between two overlapping circulo-lines $i$ and $j$ to the center of $j$. The effective contact point $C_{ij}$ is the midpoint of $r_{ij}$ as shown in Fig.~\ref{fgr:Shape} (e). We define the pressure as $P=(\Sigma_{xx}+\Sigma_{yy})/2$ and the shear stress as $\Sigma =-\Sigma_{xy}$. 

To calculate the shear modulus, we first apply an affine simple shear strain step $\delta \gamma$, which changes the circulo-line center of mass positions and orientations,
\begin{equation}
\label{affinexy}
x'_i = x_i + \delta \gamma y_i,
\end{equation}
and
\begin{equation}
\label{affineth}
    \theta'_i = \cot^{-1}(\cot \theta_i+\delta\gamma),
\end{equation}
together with Lees-Edwards periodic boundary conditions. 
($x_i$,$y_i$) gives the position of the center of mass of the $i$th circulo-line and $\theta_i$ gives the angle that the axis of the circulo-line makes with the $x$-axis. After each simple shear strain step, we minimize the total potential energy using the FIRE algorithm and measure the shear stress $\Sigma$.  We can then determine the shear modulus by calculating $G=d\Sigma/d\gamma$.

We employ an athermal, quasistatic isotropic compression protocol to generate packings of circulo-lines at jamming onset. We initialize the system with random positions and orientations of the circulo-lines in the dilute limit with packing fraction $\phi < 10^{-2}$. We then compress the system by $\Delta \phi/\phi = 2 \times 10^{-3}$, minimize the total potential energy using the FIRE algorithm, and measure the pressure $P$.  If $P < P_t$, where $P_t$ is the target pressure, we again compress the system by $\Delta \phi$ and minimize the total potential energy.  If $P > P_t$, we return to the previous configuration and decrease the packing fraction increment by a factor of two. We continue this compression process until $|P-P_t|/P_t < 10^{-5}$.  For $P_t < 10^{-7}$, we find that the number of interparticle contacts satisfies the effective isostatic condition: $N_c=N_c^0-N_q$, where $N_c^0=d_fN'-d+1$, $d=2$ is the spatial dimension, the number of degrees of freedom per particle, $d_f =3$, $N' = N - N_r - N_s/3$, $N_r$ is the number of rattler particles with too few contacts to constrain the $d_f$ degrees of freedom, $N_s$ is the number of slider particles with one unconstrained translational degree of freedom, and $N_q$ is the number of quartic modes of the dynamical matrix~\cite{Kyle_nonsphere}. To investigate the pressure-dependent mechanical properties, we start with systems at $P_t =10^{-7}$ and then successively increase the target pressure over the range $10^{-7} < P_t < 10^{-2}$.

\begin{figure}[h!]
\centering
\includegraphics[width=\linewidth,keepaspectratio]{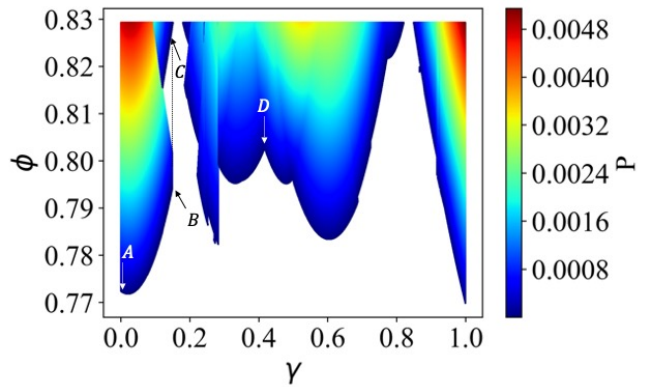}
\caption{Pressure $P$ for jammed packings of $N=6$ bidisperse disks as a function of packing fraction $\phi$ and shear strain $\gamma$. White regions correspond to unjammed states. Points A and B correspond to the beginning and end of a geometrical family in the $P \rightarrow 0$ limit. At point B ($\gamma \sim 0.17$), the system undergoes a jump change to the next geometrical family at point C. Point D indicates a point change from one geometrical family to another.}
\label{fgr:Colormap_disk}
\end{figure}

\begin{figure*}
\centering
\includegraphics[width=\textwidth,keepaspectratio]{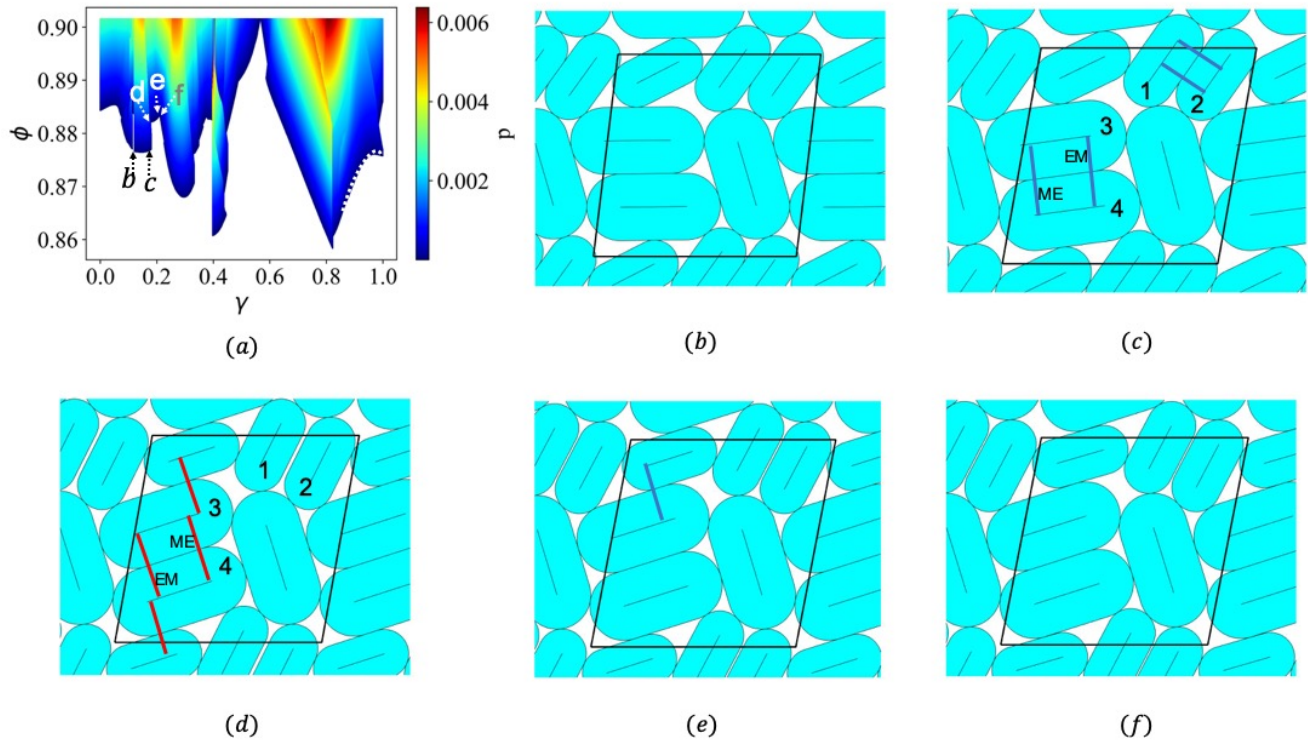}
\caption{(a) Pressure $P$ for jammed packings of $N=6$ bidisperse circulo-lines with aspect ratio ${\cal R} = 2.0$ as a function of packing fraction $\phi$ and shear strain $\gamma$. White regions correspond to unjammed states.  Points (b) and (c) correspond to the beginning and end of an upward geometrical family. At point (c) ($\gamma \sim 0.19$), the system undergoes a jump change to point (d). A second upward geometrical family occurs between points (d) and (e). The system undergoes a point change from points (e) to (f) ($\gamma \sim 0.205$). We also show a downward geometrical family in the $P\rightarrow 0$ limit (dotted line) that extends from $\gamma \sim 0.82$ to $1$. The packings in panels (b)-(f) correspond to points (b)-(f) in the $\phi$-$\gamma$ plane in panel (a). The highlighted contacts in (c) do not occur in (d), and the highlighted contacts in (d) do not occur in (c). The highlighted contact in (e) does not occur in (f). EM (or ME) indicates an end-middle (or middle-end) contact between circulo-lines $3$ and $4$.}
\label{fgr:Colormap_circulo}
\end{figure*}

\section{Results}
\label{results}

Here, we describe our main results in five subsections.  In Sec.~\ref{geometrical_families}, we generalize the concept of geometrical families of jammed packings in the $\phi$-$\gamma$ plane to packings of circulo-lines and show that the curvature $d^2\phi/d\gamma^2$ can be both positive and negative for packings of circulo-lines. In Sec.~\ref{sec:dilatancy}, we derive a general expression for the pressure-dependent shear modulus $G$ in terms of derivatives of $U$ and $\phi$ with respect to $\gamma$ at fixed packing fraction and at fixed pressure.  Using this expression, we find that the first geometrical family in the $P \rightarrow 0$ limit scales as $G_f/G_0 \sim 1 \pm P/P_0$ to linear order in $P$, where the sign of $d^2\phi/d\gamma^2$ determines the sign of the second term in $G_f$. In Sec.~\ref{nonaffine}, we decompose the shear modulus for each low-pressure geometrical family, $G_f = G_a - G_{na}$, into the affine and non-affine contributions, respectively. We show that the non-affine contribution to the shear modulus of the low-pressure geometrical families is larger for packings of circulo-lines compared to that for disk packings, and that the pressure dependence of $G_{na}$ can cause $G_f$ to increase with pressure, which does not occur for packings of spherical particles.  In Sec.~\ref{changes}, we characterize point and jump changes in the contact network during isotropic compression in packings of circulo-lines (interacting via repulsive linear springs) and show that jump changes give rise to discontinuous changes in potential energy, shear stress, and shear modulus, whereas point changes give rise to discontinuous changes only in the shear modulus. We find that point changes are more frequent for packings of circulo-lines (compared to disk packings), but the resulting jumps in the shear modulus are smaller, over the same range of pressure as for disk packings.  In Sec.~\ref{ensemble_average}, we discuss the results for the ensemble-averaged shear modulus $\langle G\rangle$ as a function of pressure.  We show that, for large aspect ratios ${\cal R} \gtrsim 1.2$, $\langle G\rangle$ scales as a power-law at large pressures with an exponent that is nearly a factor of $2$ larger than that for disk packings.  For small aspect ratios, ${\cal R} \lesssim 1.2$, $\langle G\rangle$ versus pressure does not possess a single power-law scaling exponent.  
We further decompose $\langle G\rangle = \langle G_f \rangle + \langle G_r\rangle$ into contributions from geometrical families $G_f$ and changes in the contact network $G_r$. We show that $\langle G_r \rangle$ is smaller for packings of circulo-lines compared to that for spherical particles, and thus the increase in the power-law exponent is caused by the fact that the shear modulus can increase in pressure during geometrical families for packings of circulo-lines. 

\subsection{Geometrical families of circulo-lines}
\label{geometrical_families}

Jammed packings within a geometrical family are mechanically stable packings with different values of the pressure (and shear stress) that are related to each other by continuous, quasistatic changes in packing fraction and shear strain with no changes in the interparticle contact network. We showed in previous studies of jammed disk packings that  near-isostatic geometrical families form upward parabolic segments in the $\phi$-$\gamma$ plane~\cite{protocol,Sheng_stress_anisotropy}.  An example geometrical family of $N=6$ jammed disk packings in the $P \rightarrow 0$ limit runs from point A to B in Fig.~\ref{fgr:Colormap_disk}.  Similarly shaped geometrical families occur at higher pressures, as shown by the upward parabolic contours in Fig.~\ref{fgr:Colormap_disk}. 

Geometrical families begin and end at point or jump changes in the interparticle contact network~\cite{PJ_jump_changes}. For example, after making an infinitesimal increase in shear strain at point B ($\gamma \sim 0.17$) in Fig.~\ref{fgr:Colormap_disk}, the system undergoes a jump change to point C. At a jump change, the packing becomes unstable, particles rearrange, and the packing fraction, shear stress, and shear modulus change discontinuously. At point D ($\gamma \sim 0.40$) in Fig.~\ref{fgr:Colormap_disk}, the system undergoes a point change. For point changes, a single contact is added or removed from the contact network and the packing fraction and shear stress are continuous.  For packings with purely repulsive linear spring interactions, the shear modulus is discontinuous at point changes~\cite{PJ_jump_changes}.   

\begin{figure}[h!]
\centering
\includegraphics[width=8cm,keepaspectratio]{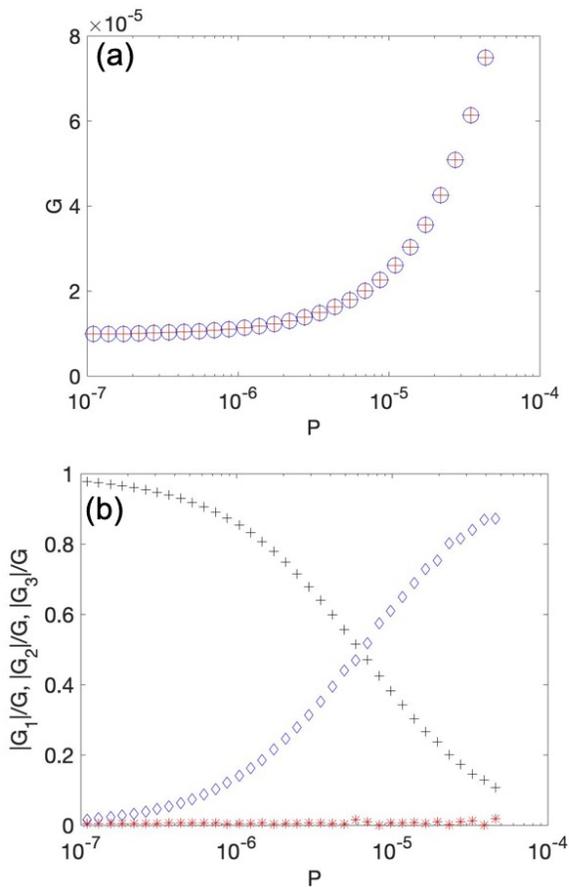}
\caption{(a) Comparison of the shear modulus $G$ versus pressure $P$ for $N=16$ jammed packings of bidisperse circulo-lines with ${\cal R}=2$ obtained by calculating $G=d\Sigma/d\gamma$ numerically (circles) and by calculating the three terms $G=G_1+G_2+G_3$ in Eq.~\ref{total} individually (crosses). Note that $G$ increases with $P$ for this geometrical family. (b) $|G_1|/G$ (asterisks), $|G_2|/G$ (diamonds), and $|G_3|/G$ (crosses) as defined in Eq.~\ref{total} for the same data in (a).}
\label{fgr:3_term}
\end{figure}

In Fig.~\ref{fgr:Colormap_circulo} (a), we show the structure of geometrical families in the $\phi$-$\gamma$ plane for jammed packings of $N=6$ bidisperse circulo-lines with aspect ratio ${\cal R}=2.0$.  As for jammed disk packings, we find that packings of circulo-lines can form upward parabolic geometrical families, e.g. from point (b) to (c) in panel (a).  The configurations that correspond to the beginning and end of this geometrical family are shown in Fig.~\ref{fgr:Colormap_circulo} (b) and (c). These two configurations share the same contact network, but the relative angles that the  circulo-lines make with each other are different, e.g. circulo-lines $3$ and $4$ in panel (c) are tilted away from the horizontal axis compared to those in panel (b). 

Because pairs of circulo-lines can form three different types of contacts, the contact network can change when particles $i$ and $j$ form or break a contact, as well as when the type of contact between $i$ and $j$ changes (e.g. from an end-end to an end-middle contact). (See Sec.~\ref{methods}.)  As a result, packings of circulo-lines possess more point changes and shorter geometrical families in the $\phi$-$\gamma$ plane than disk packings. For example, the average terminus $P_{\rm end}$ of the lowest-pressure geometrical family is a factor of $5$ larger for $N=6$ disk packings compared to $N=6$ packings of circulo-lines with ${\cal R}=2$.  In response to an infinitesimal increase in shear strain at point (c) in Fig.~\ref{fgr:Colormap_circulo} (a), the system undergoes a jump change to point (d).  There are a total of eight changes in the contact network across the jump change, e.g. circulo-lines $1$ and $2$ are in contact in Fig.~\ref{fgr:Colormap_circulo} (c), but they are not in contact in (d) and the types of contacts between circulo-lines $3$ and $4$ change between Fig.~\ref{fgr:Colormap_circulo} (c) and (d) (from end-middle to middle-end or from middle-end to end-middle). At a jump change in packings of circulo-lines, the packing fraction, shear stress, and shear modulus also change discontinuously.  (See Sec.~\ref{changes} below.)

Another upward geometrical family occurs between points (d) and (e) in 
Fig.~\ref{fgr:Colormap_circulo} (a). At point (e), the system undergoes a point change. As shown in Fig.~\ref{fgr:Colormap_circulo} (e) and (f), the system loses a single contact across the point change. The packing fraction and shear stress are continuous, whereas we will show in Sec.~\ref{changes} that the shear modulus is discontinuous across a point change (for purely repulsive, linear spring interactions).

In contrast to jammed disk packings, the geometrical families for packings of circulo-lines can form both upward and downward parabolic segments in the $\phi$-$\gamma$ plane.  One of the downward geometrical families is highlighted from $\gamma \sim 0.82$ to $1$ in Fig.~\ref{fgr:Colormap_circulo} (a).  In the next section, we show that the sign of the curvature of the parabolic geometrical families determines whether the shear modulus of the geometrical family increases or decreases with pressure. 

\begin{figure}[h!]
\centering
\includegraphics[height=6cm]{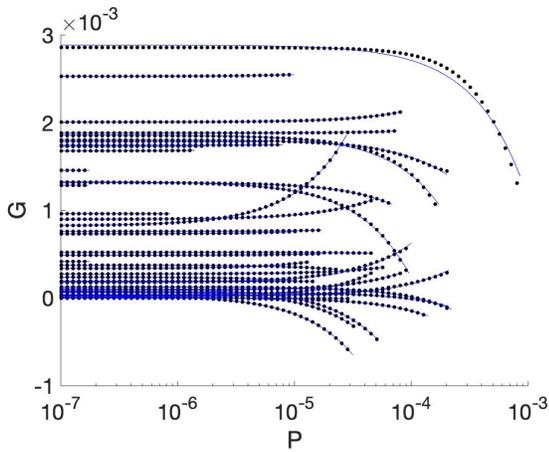}
\caption{The shear modulus $G$ versus the pressure $P$ for $N=16$ jammed packings of bidisperse circulo-lines with aspect ratio ${\cal R} =2.0$.  We show $50$ different low-pressure geometrical families (black circles) and each geometrical family ends at a point or jump change in the interparticle contact network. The blue solid lines are best fits to $G=G_0 + \eta P$.}
\label{fgr:single_family}
\end{figure}

\subsection{Stress-dilatancy relation}
\label{sec:dilatancy}

We now derive an exact expression for the shear modulus in terms of derivatives of the total potential energy $U$, packing fraction $\phi$, and pressure $P$ with respect to the shear strain $\gamma$. This expression will enable us to understand the behavior of $G(P)$ within geometrical families. If we take infinitesmal steps $d\phi$ and $d\gamma$ along a geometrical family in the $\phi$-$\gamma$ plane, the change in the total potential energy is 
\begin{equation}
\label{differential}
    dU = -PdA - \Sigma_{xy}Ad\gamma,
\end{equation}
where $A=L^2$ is the area of the system and $dA/A = -d\phi/\phi$.  After rearranging Eq.~\ref{differential}, we find the following expression that relates the shear stress to the dilatancy $-\phi^{-1} (d\phi/d\gamma)_{P}$ at finite pressure~\cite{Sheng_stress_anisotropy}:
\begin{equation}
\label{dilatancy}
    \Sigma = \frac{1}{L^2} \left(\frac{dU}{d\gamma}\right)_P - \frac{P}{\phi} \left(\frac{d\phi}{d\gamma}\right)_P.
\end{equation}
The shear modulus can be obtained by calculating the derivative $d \Sigma/d\gamma$ at constant packing fraction:
\begin{equation}
\label{total}
\begin{aligned}
    G = \left( \frac{d \Sigma}{d\gamma} \right)_{\phi} =\frac{1}{L^2}\left(\frac{d\left(\frac{dU}{d\gamma}\right)_{P}}{d\gamma} \right)_\phi \\ - \frac{P}{\phi}\left(\frac{\left(\frac{d\phi}{d\gamma}\right)_P}{d\gamma} \right)_\phi - \frac{1}{\phi}\left(\frac{dP}{d\gamma}\right)_\phi\left(\frac{d\phi}{d\gamma} \right)_P. 
    \end{aligned}
\end{equation}
The shear modulus is a sum of three terms, $G=G_1+G_2+G_3$, where $G_1=L^{-2} (d (dU/d\gamma)_P/d\gamma)_{\phi}$ includes mixed derivatives of $U$ with respect to $\gamma$ at fixed $\phi$ and at fixed $P$, $G_2=-P\phi^{-1} (d (d\phi/d\gamma)_{P}/d\gamma)_{\phi}$ is proportional to the derivative of the dilatancy with respect to $\gamma$ at fixed $\phi$, and $G_3=-\phi^{-1} (dP/d\gamma)_{\phi} (d\phi/d\gamma)_{P}$ includes shear strain derivatives of $P$ at fixed $\phi$ and of $\phi$ at fixed $P$. Eq.~\ref{total} is verified numerically for a low-pressure geometrical family of circulo-line packings for which $G$ increases with $P$ in Fig.~\ref{fgr:3_term} (a).  For all disk and circulo-line packings that we have considered, $|G_1| \ll |G_2|$ or $|G_1| \ll |G_3|$, and thus the first term $G_1$ in Eq.~\ref{total} can be neglected.  Further, we show in Fig.~\ref{fgr:3_term} (b) for an $N=16$ packing of circulo-lines with ${\cal R}=2$ that $|G_3| > |G_2|$ for $P < P'$ and $|G_2| > |G_3|$ for $P > P'$.  ($P' \sim 10^{-5}$ for this particular system size and aspect ratio.) We find that $G \sim G_3 \sim G_0>0$ in the zero-pressure limit, and thus the shear modulus for low-pressure geometrical families can be approximated as $G \sim G_0 + \eta P$, where $\eta$ is determined by the negative curvature of the geometrical families in the $\phi$-$\gamma$ plane. In particular, $\eta >0$ ($\eta <0$) for downward (upward) geometrical families in the $\phi$-$\gamma$ plane. In Fig.~\ref{fgr:single_family}, we show the shear modulus versus pressure for $50$ low-pressure geometrical families of $N=16$ bidisperse packings of circulo-lines with ${\cal R}=2$. We show that the shear modulus obeys $G \sim G_0 + \eta P$ within each geometrical family. For this system size and aspect ratio, approximately half of the geometrical families have $\eta > 0$ and the other half have $\eta <0$. The fact that a finite fraction of geometrical families possesses upward geometrical families with $\eta >0$ strongly influences the ensemble-averaged shear modulus $\langle G\rangle$ as discussed in Sec.~\ref{ensemble_average}. 

\subsection{Affine and nonaffine contributions to the shear modulus for geometrical families}
\label{nonaffine}

The shear modulus can be decomposed into the affine and non-affine contributions: $G=G_a-G_{na}$, where the affine contribution, $G_a=L^{-2} d^2U/d\gamma^2$, is obtained by applying a global, affine simple shear strain and the non-affine contribution, $G_{na}$, includes the relaxation process during potential energy minimization following the applied affine shear strain. Numerous prior studies have shown that the non-affine response dominates the shear modulus in disk packings near jamming onset ~\cite{mizuno2016elastic,ellenbroek2009non,schlegel2016local,zaccone2011approximate,phillips2021universal}.  Similar calculations of the non-affine contribution to the shear modulus of jammed packings of non-spherical particles have not been performed.  

\begin{figure}[h!]
\centering
\includegraphics[height=14cm]{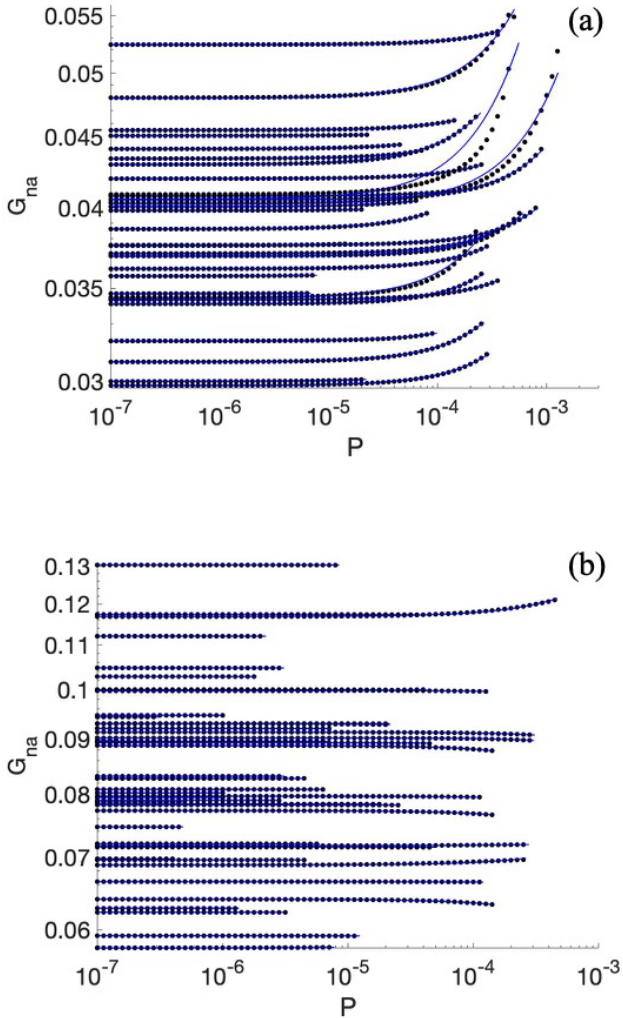}
\caption{(a) The non-affine contribution to the shear modulus $G_{na}$ versus pressure $P$ for $30$ low-pressure geometrical families of $N=16$ jammed disk packings (black dots). (b) $G_{na}$ versus $P$ for $57$ low-pressure geometrical families of $N=16$ bidisperse circulo-line packings with ${\cal R}=2.0$ (black dots). Each geometrical family ends at a point or jump change in the interparticle contact network. The solid blue lines in both panels are fits to $G_{na} \sim G'_{0} + \eta' P$.}
\label{fgr:Gna}
\end{figure}

In previous studies~\cite{Philip_Herzian}, we calculated $G_{na} = G - G_a$ for isostatic geometrical families of jammed disk packings.  These results are also shown in Fig.~\ref{fgr:Gna} (a) for $N=16$ disk packings. We find that $G_{na}$ increases with pressure, which causes jammed disk packings to soften (i.e. $G$ decreases) along compression geometrical families.  When the geometrical families persist to high pressure, we find that $G_{na}$ increases rapidly near point and jump changes in the contact network, which causes deviations from the simple form $G_{na} \sim G_0' + \eta' P$. 

We now describe similar studies of $G_{na}$ for low-pressure geometrical families of circulo-lines, as shown in Fig.~\ref{fgr:Gna} (b) for ${\cal R}=2.0$.  (We show explicit expressions for the affine shear modulus $G_a$ for packings of circulo-lines in Appendix~\ref{appA}.) We identify several results concerning $G_{na}$ for packings of circulo-lines that are different from the results for disk packings. First, as discussed above, the average pressure range over which the first, low-pressure geometrical families persist is smaller for packings of circulo-lines, and $G_{na}(P)$ is well-fit by $G_{na}(P) \sim G_0' +\eta' P$.  Second, $G_{na}$ can either increase or decrease with pressure (i.e. $\eta' >0$ or $\eta' <0$).  In particular, packings of circulo-lines can harden (i.e. $G$ increases) along compression geometrical families, in contrast to disk packings.  In addition, the ensemble-averaged $\langle G_{na} \rangle$ for the first geometrical family is typically larger for packings of circulo-lines compared to that for disk packings.  For example, for packings of circulo-lines with ${\cal R}=2.0$, $\langle G_{na}\rangle$ is more than a factor of $2$ larger than $\langle G_{na}\rangle$ for disk packings.  

\subsection{Point and jump changes in the contact network}
\label{changes}

\begin{figure}[h!]
\centering
\includegraphics[height=13cm]{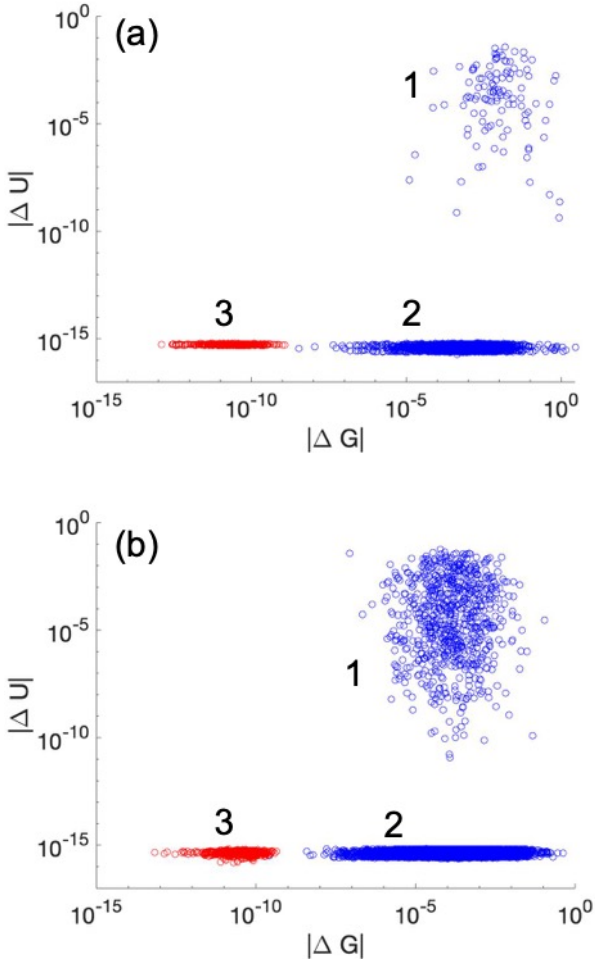}
\caption{(a) Scatter plot of the change in the shear modulus $|\Delta G|$ versus the change in the potential energy $|\Delta U|$ measured between packings separated by small compression steps for $N=128$ disk packings. (b) Same data as in (a), but for $N=128$ circulo-line packings with ${\cal R} =2.0$. The red dots in region $3$ indicate comparisons between two packings with the same interparticle contact networks. The blue points in regions $1$ and $2$ correspond to jump and point changes, respectively.}
\label{fgr:change}
\end{figure}

In previous studies~\cite{PJ_jump_changes} of jammed packings of disks that interact via purely repulsive, linear spring potentials, we found that the shear modulus possesses discontinuous jumps when the packings undergo point and jump changes in the interparticle contact networks during applied deformations (such as shear or compression). Point changes are additions or removals of a single contact in the network. In contrast, jump changes are caused by mechanical instabilities and involve multiple contact changes and significant particle motion. In Fig.~\ref{fgr:change} (a), we show a scatter plot of changes in $G$ and the potential energy $U$ between jammed $N=128$ disk packings that are separated by small compression steps over a range of pressure $10^{-3} < P < 10^{-2}$.  To generate Fig.~\ref{fgr:change} (a), we identify the pressures at the beginning and end of all geometrical families in this pressure range ($10^{-3} < P < 10^{-2}$) and compare $G$ and $U$ between two packings before the contact network change and one before and one after the contact network change. 

We identify three regions of clustered points in Fig.~\ref{fgr:change} (a): regions with 1) large $|\Delta G|$ and large $|\Delta U|$, 2) large $|\Delta G|$ and small $|\Delta U|$, or 3) small $|\Delta G|$ and small $|\Delta U|$.  The interparticle contact networks change for the systems in regions $1$ and $2$, and data in these regions correspond to jump and point changes, respectively.  For the data in region $3$, the contact network does not change, and $|\Delta G|$ and $|\Delta U|$ will decrease with improved force balance. 

In Fig.~\ref{fgr:change} (b), we show similar data for $|\Delta G|$ and $|\Delta U|$ for $N=128$ jammed packings of circulo-lines with ${\cal R} =2.0$ over the same pressure range as studied for jammed disk packings. As found previously for disk packings, jammed packings of circulo-lines undergo point and jump changes in the contact network, which lead to discontinuous jumps in $G$.  As found previously, point and jump changes can be differentiated because point changes have vanishing $|\Delta U|$, whereas jump changes have non-zero values of $|\Delta U|$. 

For the jammed disk packings considered in Fig.~\ref{fgr:change} (a), jump changes accounted for $\sim 0.13$ of the changes in the contact network.  However, jump changes accounted for only $\sim 0.03$ of the changes in the contact network for jammed circulo-line packings over the same range of pressure.  Since point changes can involve changes in the {\it type} of contact between the same pair of particles, point changes are much more frequent in packings of circulo-lines compared to disk packings.  

\subsection{Ensemble-averaged shear modulus}
\label{ensemble_average}

\begin{figure}[h!]
\centering
\includegraphics[height=14cm]{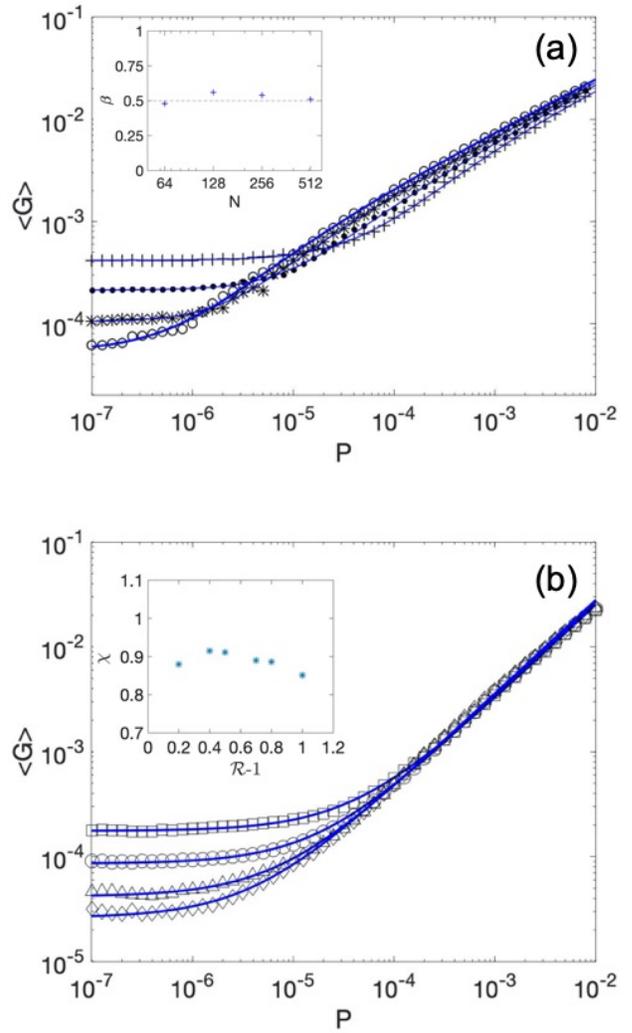}
\caption{(a) Ensemble-averaged shear modulus $\langle G\rangle$ versus pressure $P$ for bidisperse disk packings for several system sizes: $N=64$ (crosses), $128$ (filled circles), $256$ (asterisks), and $512$ (open circles). The solid lines are best fits to Eq.~\ref{pressure}. The inset shows the large-pressure power-law scaling exponent $\beta$ versus $N$. (b) Similar data as in (a), but for packings of circulo-lines with ${\cal R} = 1.5$ and $N=64$ (squares), $128$ (circles), $256$ (triangles), and $512$ (diamonds). The solid lines are best fits to Eq.~\ref{pressure} with $c=0$. The inset shows the large-pressure scaling exponent $\chi$ versus ${\cal R}-1$. In both panels, the averages are calculated over $10^3$ different initial conditions.}
\label{fgr:Gsys}
\end{figure}

\begin{figure}[h!]
\centering
\includegraphics[height=14cm]{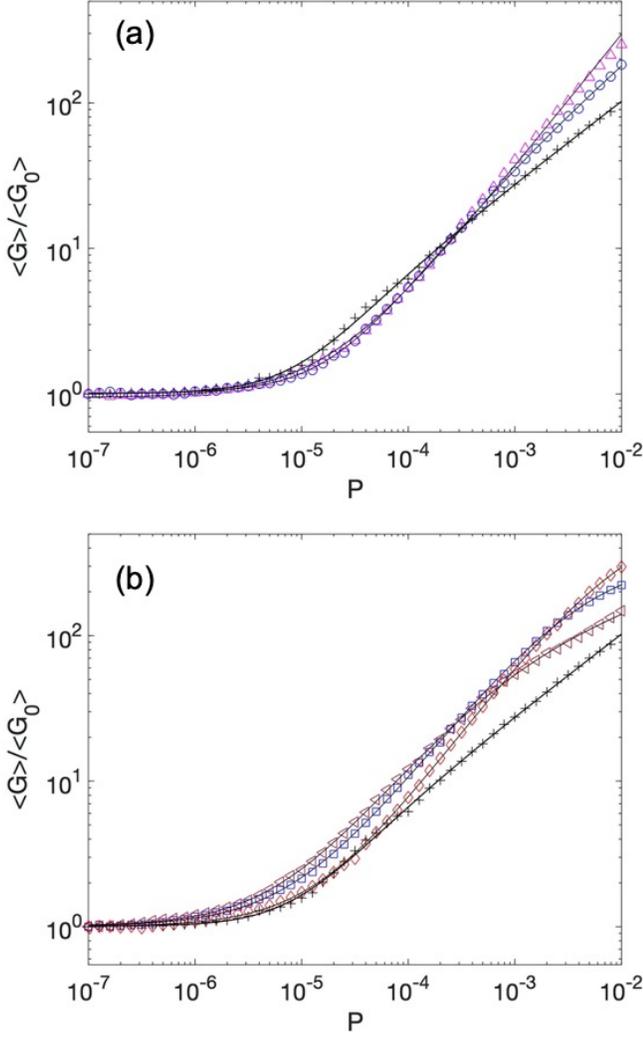}
\caption{(a) Ensemble-averaged shear modulus $\langle G\rangle$ (normalized by the zero-pressure value $\langle G_0\rangle$) versus pressure $P$ for $N=128$ packings of circulo-lines with several aspect ratios: ${\cal R}=1$ (crosses), $1.5$ (upward triangles), and $2$ (circles).  (b) Similar data in (a), except for ${\cal R}=1$ (crosses), $1.05$ (leftward triangles), $1.1$ (squares), and $1.2$ (diamonds).  In both panels, the solid lines are fits to Eq.~\ref{pressure}.  For ${\cal R} > 1.2$, $c=0$. $\langle G \rangle$ is calculated using $10^3$ different initial conditions.}
\label{fgr:Gasp}
\end{figure}

\begin{figure}[h!]
\centering
\includegraphics[height=14cm]{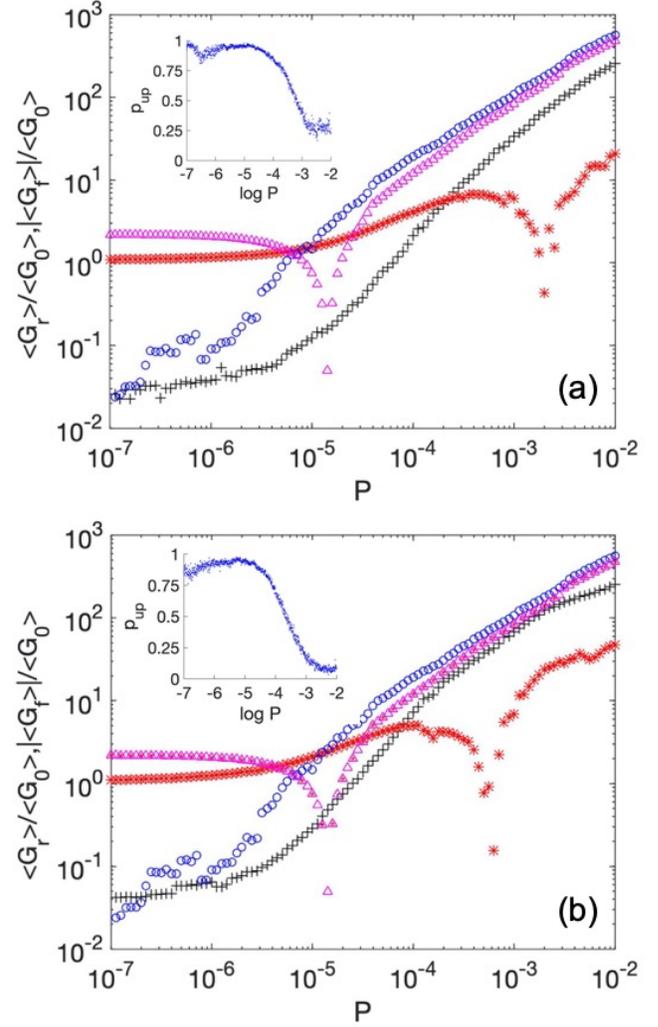}
\caption{The average rearrangement $\langle G_r \rangle$ (crosses) and geometrical family $|\langle G_f \rangle|$ (asterisks) contributions to the shear modulus for $N=128$ jammed packings of criculo-lines (normalized by the average zero-pressure value of the shear modulus, $\langle G_0\rangle$) for (a) ${\cal R} =1.5$ and (b) $1.1$. As a comparison, we also show $\langle G_r \rangle$ (circles) and $|\langle G_f \rangle|$ (triangles) for $N=128$ disk packings. In both panels, the inset shows the probability $p_{\rm up}$ that each geometrical family increases with pressure, obtained from $500$ jammed packings of circulo-lines at each small pressure interval.}
\label{fgr:GfGr}
\end{figure}

Numerous prior studies have shown that the ensemble-averaged shear modulus $\langle G \rangle$ for packings of frictionless, spherical particles increases as a power-law in pressure when $P > P^{**}$~\cite{goodrich,epitome}, where $P^{**} \sim N^{-2}$ decreases with increasing system size.  For finite-sized systems, we have used the following scaling form for the ensemble-averaged shear modulus~\cite{Kyle_PRL}:
\begin{equation}
    \label{pressure}
    \langle G \rangle = \langle G_0 \rangle + \frac{b P^{\chi}}{1 + cP^{\chi-\beta}},  
\end{equation}
where $\langle G_0 \rangle \sim N^{-1}$, $b$ and $c$ are constants, $\chi$ and $\beta$ are power-law exponents, and the large-pressure scaling exponent $\beta \sim 0.5$ for packings of spherical particles with repulsive linear spring interactions.  (See Fig.~\ref{fgr:Gsys} (a).)

In Fig.~\ref{fgr:Gsys} (b), we show $\langle G\rangle$ versus $P$ for packings of circulo-lines with ${\cal R} = 1.5$ for several system sizes.  As for disk packings, $\langle G \rangle = \langle G_0\rangle \sim N^{-1}$ in the zero-pressure limit and $\langle G(P) \rangle$ increases as a power-law at large pressure.  For packings of circulo-lines with ${\cal R} \gtrsim 1.2$, we find that a scaling function with a single power-law (i.e. $c=0$ in Eq.~\ref{pressure}) provides a better description of $\langle G \rangle$ versus $P$ over the full range of pressure. We find that the power-law exponent $\chi \sim 0.9$ for ${\cal R} =1.5$, suggesting that the pressure-dependent mechanical properties of jammed packings of circulo-lines differ from those of jammed disk packings.    

In Fig.~\ref{fgr:Gasp}, we show $\langle G \rangle$ (normalized by the zero-pressure value $\langle G_0\rangle$) versus $P$ over a range of aspect ratios (for a single system size $N=128$).  In panel (a), we compare $\langle G\rangle/\langle G_0\rangle$ for packings of circulo-lines with large aspect ratios (${\cal R} =1.5$ and $2$) and for disk packings.  $\langle G \rangle/\langle G_0\rangle$ for the packings with large aspect ratios has a robust large-pressure scaling exponent $\chi$ that is larger than that for disk packings. The inset to Fig.~\ref{fgr:Gsys} (b) shows that $0.8 \lesssim \chi \lesssim 0.9$ for ${\cal R} > 1.2$. 
In panel (b), we show $\langle G\rangle/\langle G_0\rangle$ versus $P$ for small aspect ratios ${\cal R} \le 1.2$.  $\langle G\rangle/\langle G_0\rangle$ no longer has a single large-pressure scaling exponent. The curves have a steep region in the intermediate pressure regime from $10^{-5}$ to $10^{-3.5}$, with scaling exponents that are comparable to those for higher aspect ratios.  However, at larger pressures above a characteristic pressure, $P>P^*$, the curves bend over and have scaling exponents that are much less than those in the inset to Fig.~\ref{fgr:Gsys} (b). The data also suggests that $P^*$ decreases as the aspect ratio decreases, indicating that the range of pressure over which the elevated scaling exponent occurs decreases as ${\cal R} \rightarrow 1$.

To explain the increase in the power-law scaling exponent for packings of circulo-lines, we decompose the ensemble-averaged shear modulus $\langle G\rangle$ into contributions from geometrical families, $\langle G_f \rangle$, and from discontinuous jumps caused by changes in the interparticle contact network, $\langle G_r\rangle$: $\langle G \rangle = \langle G_f \rangle + \langle G_r \rangle$~\cite{Kyle_PRL,Philip_Herzian}. In Fig.~\ref{fgr:GfGr} (a), we compare $|\langle G_f\rangle|/\langle G_0\rangle$ and $\langle G_r \rangle/\langle G_0 \rangle$ for packings of circulo-lines with ${\cal R} = 1.5$ and for disk packings both with $N=128$. For jammed disk packings, $\langle G_f \rangle/\langle G_0\rangle$ decreases monotonically with increasing pressure, and thus there is a characteristic pressure $\langle P_0\rangle$ at which $\langle G_f \rangle = 0$ and above which $\langle G_f \rangle <0$.  For $N=128$ jammed disk packings, $\langle P_0 \rangle \sim 10^{-5}$.  Thus, for $P > \langle P_0 \rangle$, the difference between the rearrangement and geometrical family contributions, $\langle G\rangle = \langle G_r\rangle - |\langle G_f\rangle|$, determines the power-law scaling behavior of the shear modulus with pressure for jammed disk packings.  

A key difference between jammed disk packings and packings of circulo-lines is that circulo-line packings possess a finite fraction of geometrical families whose shear modulus increases with pressure (see inset of Fig.~\ref{fgr:GfGr} (a)), which can cause $\langle G_f \rangle/\langle G_0\rangle$ to {\it increase} with pressure.  For ${\cal R} =1.5$ circulo-line packings with $N=128$, we find that $\langle G_f\rangle/\langle G_0\rangle$ increases over a wide pressure range from $10^{-7} \lesssim P \lesssim 10^{-3.5}$. For $P \gtrsim 10^{-3.5}$, the fraction of geometrical families that increases with pressure, $p_{\rm up}$, decreases dramatically, and $\langle G_f\rangle/\langle G_0\rangle$ begins decreasing with pressure. $\langle G_f\rangle/\langle G_0\rangle$ reaches zero near $\langle P_0 \rangle\sim 10^{-2.5}$ and continues decreasing with further increases in pressure.  

In addition, the rearrangement contribution $\langle G_r\rangle/\langle G_0\rangle$ is larger for disk packings compared to packings of circulo-lines with ${\cal R}=1.5$, as shown in Fig.~\ref{fgr:GfGr} (a).  Thus, even though the frequency of contact network changes is enhanced for packings of circulo-lines (see Sec.~\ref{changes}), the discontinuous jumps in the shear modulus are sufficiently small for circulo-line packings that $\langle G_r\rangle/\langle G_0\rangle$ is larger for disk packings.  Since $\langle G_r \rangle/\langle G_0\rangle$ is larger for jammed disk packings compared to circulo-line packings and $\langle G_f\rangle/\langle G_0\rangle$ increases with pressure over a wide range of pressure, it is clear that the existence of geometrical families with shear moduli that increase with pressure is responsible for the elevanted power-law scaling exponent for packings of circulo-lines with ${\cal R} \gtrsim 1.2$.

For jammed packings of circulo-lines with ${\cal R} \lesssim 1.2$, we do not find a single power-law scaling exponent for $\langle G\rangle$ versus $P$.  Instead, $\langle G(P)\rangle$
has a power-law exponent $\beta \sim 0.8$-$0.9$ for intermediate pressures and then the exponent decreases for $P \gtrsim 10^{-2.5}$.  (See Fig.~\ref{fgr:Gasp} (b).) In Fig.~\ref{fgr:GfGr} (b), we show that $\langle G_r \rangle$ is much larger for circulo-line packings with ${\cal R} =1.1$ than that with $1.5$. In particular, $\langle G_r \rangle$ for ${\cal R}=1.1$ is comparable to that for jammed disk packings for $P \gtrsim 10^{-4}$ and $\langle G_f\rangle > 0$ over a much narrower range of pressure.  The two results cause the lack of single power-law scaling for packings of circulo-lines with ${\cal R} \lesssim 1.2$. 

\section{Conclusions and future directions}
\label{conclusions}

In this article, we studied the structural and mechanical properties of jammed packings of circulo-lines with frictionless, purely repulsive, linear spring interactions. We found several important results for jammed packings of circulo-lines that are different from those for jammed packings of spherical particles.  First, we showed that packings of circulo-lines posses geometrical families that can be both concave upward or concave downward in the packing fraction-shear strain ($\phi$-$\gamma$) plane.  In contrast, the geometrical families are nearly always concave upward in the $\phi$-$\gamma$ plane, especially at low pressure, for jammed packings of spherical particles.  We then derived a stress-dilatancy relation for packings at finite pressure, which allowed us to show that the shear modulus for low-pressure geometrical families obeys $G_f = G_0 +\eta P$ to linear order in pressure, where the sign of $\eta$ is determined by the negative curvature of geometrical families in the $\phi$-$\gamma$ plane.  Thus, the shear modulus of low-pressure geometrical families increases with pressure when $d^2\phi/d\gamma^2 < 0$ and decreases with pressure when $d^2\phi/d\gamma^2 > 0$.  The fact that the shear modulus of geometrical families can increase with pressure has a profound effect on the pressure-dependent, ensemble-averaged shear modulus $\langle G(P)\rangle$.  In particular, we found that $\langle G(P)\rangle$ for jammed packings of circulo-lines with aspect ratios ${\cal R} \gtrsim 1.2$ displays robust power-law scaling over a wide range of pressure, but the scaling exponent ($\beta \sim 0.8$-$0.9$) is nearly a factor of two larger than that for jammed disk packings ($\beta \sim 0.5$). For smaller aspect ratios, ${\cal R} \lesssim 1.2$, $\langle G(P)\rangle$ does not possess a single power-law scaling exponent over the same range of pressure.  To understand the origin of this behavior, we decomposed $\langle G\rangle$ into separate contributions from geometrical families, $\langle G_f \rangle$, and from changes in the contact network, $\langle G_r\rangle$: $\langle G\rangle/\langle G_0\rangle = \langle G_f \rangle/\langle G_0 \rangle + \langle G_r\rangle/\langle G_0 \rangle$, where $\langle G_0 \rangle$ is the value of $\langle G\rangle$ in the zero-pressure limit.   In general, we found that $\langle G_r\rangle/\langle G_0 \rangle$ is larger for disk packings compared to that for packings of circulo-lines, even though the frequency of changes in the contact network is larger for packings of circulo-lines.  In contrast, we found that $\langle G_f\rangle/\langle G_0 \rangle$ is much larger for packings of circulo-lines.  In fact, $\langle G_f\rangle/\langle G_0 \rangle <0$ for disk packings, whereas it can be positive for packings of circulo-lines in the pressure regime where the power-law scaling exponent is larger than that for disk packings. Thus, the presence of geometrical families with shear moduli that increase with pressure gives rise to important changes in the pressure-dependent mechanical properties for jammed packings
of circulo-lines. 

These results suggest several promising areas of future research. First, in the present studies, we did not examine in detail how properties of jammed packings of circulo-lines in the ${\cal R} \rightarrow 1$ limit compare to those for jammed disk packings.  Two issues arise in the ${\cal R} \rightarrow 1$ limit. 1) Non-circular particles always possess $3$ degrees of freedom per particle, whereas smooth disks possess only $2$ nontrivial degrees of freedom per particle. Thus, in future studies, we will compare the properties of jammed packings of circulo-lines to those for packings of weakly frictional or bumpy particles, which both possess three degrees of freedom per particle~\cite{isostaticity}. 2) The current force model for circulo-lines, which considers forces between the end and middle sections of pairs of circulo-lines, does not converge to the force model obtained from Eq.~\ref{energy} in the ${\cal R}\rightarrow 1$ limit for packings with finite pressure.  A force law that is a function of the square-root of the area of overlap between pairs of circulo-lines is a more promising model.  

In addition, we know that jammed packings of circulo-lines possess concave upward and concave downward geometrical families in the $\phi$-$\gamma$ plane.  Do jammed packings of other non-spherical particle shapes also possess concave upward and concave downward geometrical families?  Can we find paticle shapes for which jammed packings only possess concave downward geometrical families in the $\phi$-$\gamma$ plane? We have shown in previous studies that the power-law scaling exponent for $\langle G(P)\rangle$ for jammed packings of ellipse-shaped particles is also elevated relative to that for jammed disk packings~\cite{mailman,Shreck_dimer_ellipse}.  Thus, 
it is likely that jammed packings of ellipse-shaped particles also possess concave downward geometrical families in the $\phi$-$\gamma$ plane.  Since the power-law scaling exponent $\beta$ for $\langle G(P)\rangle$ depends on properties of the geometrical families, the frequency of contact network changes, and the size of the discontinuous jumps in $G$ caused by the contact network changes, it seems likely that the power-law scaling exponent $\beta$ will depend sensitively on particle shape.  Thus, it will be important to study $\langle G(P)\rangle$ and other mechanical properties for packings of many different particle shapes in both two- and three-dimensions. 

\begin{acknowledgments}
We acknowledge support from the Army Research Laboratory under Grant No. W911NF-17-1-0164 (P.W., N.O., and C.O.), NSF Grants No. DBI-1755494 (P.T.), No. CBET-2002782 (CO.), and No. CBET-2002797 (M.S.), and China Scholarship Council Grant No. 201906340202 (S.Z.). This work was also supported by the High Performance Computing facilities operated by Yale’s Center for Research Computing and computing resources provided by the Army Research Laboratory Defense University Research Instrumentation Pro- gram Grant No. W911NF-18-1-0252. 
\end{acknowledgments}

\appendix{}

\section{Affine shear modulus}
\label{appA}

In Sec.~\ref{nonaffine} in the main text, we calculated the non-affine shear modulus $G_{na}=G-G_a$ for low-pressure geometrical families for jammed packings of circulo-lines. 
In this Appendix, we derive an expression for the affine contribution to the shear modulus, $G_a$, for jammed packings of circulo-lines. For a globally affine simple shear strain, the particle positions and orientations of each circulo-line change according to Eqs.~\ref{affinexy} and~\ref{affineth}, and the affine contribution to the shear stress is
\begin{equation}
\label{affine_stress}
      \Sigma_a=L^{-2} {\frac{dU(r_{ij})}{d\gamma}},
\end{equation}
where the contact distance vector satisfies
\begin{equation}
\label{rij}
      \vec{r}_{ij}=\vec{c}_{ij}+\lambda_j \hat{u}_j-\lambda_i \hat{u}_i,
\end{equation}
$\vec{c}_{ij}$=($x_{ij}$,$y_{ij}$) is the the separation vector between the centers of mass of particles $i$ and $j$, and $\hat{u}_i=(\cos \theta_i,\sin \theta_i)$. We set $\lambda_i=l_i$ and $\lambda_j=l_j$ for end-end contacts and set  
\begin{equation}
\label{lambda2}
      \lambda_i=(\vec{r}_{ij}+\lambda_j \hat{u}_j) \cdot \hat{u}_i
\end{equation}
for end-middle contacts, where $\lambda_j=l_j$ and particle $j$ is the particle whose end is in contact with the middle section of particle $i$. 

The affine shear modulus is 
\begin{equation}
\label{affine_modulus}
      G_a=\frac{d\Sigma_a}{d\gamma}=L^{-2} \frac{d^2U(r_{ij})}{d\gamma^2}.
\end{equation}
Using Eq.~\ref{energy}, we obtain the following for the second derivative of the total potential energy with respect to shear strain $\gamma$:
\begin{equation}
    \frac{d^2U(r_{ij})}{d\gamma^2}=\frac{1}{\sigma_{ij}^2}\left( \left(\frac{dr_{ij}}{d\gamma}\right)^2-(\sigma_{ij}-r_{ij})\frac{d^2r_{ij}}{d\gamma^2}\right).
\end{equation}
For both end-end and end-middle contacts, the first- and second-derivatives of the contact distance with strain are given by   
\begin{equation}
\label{drijdgam}
      \frac{dr_{ij}}{d\gamma}=\frac{f_1 x_{ij}+f_2 y_{ij}}{r_{ij}}
\end{equation}
and
\begin{equation}
\label{d2rijdam2}
      \frac{d^2r_{ij}}{d\gamma^2}=-\frac{(f_1 x_{ij}+f_2 y_{ij})^2}{r^3_{ij}} + \frac{f_1^2+f_2^2+f_3 x_{ij}+f_4 y_{ij}}{r_{ij}},
\end{equation}
respectively, where $f_1$, $f_2$, $f_3$, and $f_4$ are functions of $\lambda_i$, $\lambda_j$, $\theta_i$, and $\theta_j$. For end-end contacts,
\begin{equation}
\label{f1}
      f_1=y_{ij}-\lambda_i \sin^3 \theta_i+\lambda_j \sin^3 \theta_j,
\end{equation}

\begin{equation}
\label{f2}
      f_2=\lambda_i \sin^2 \theta_i \cos \theta_i-\lambda_j \sin^2 \theta_j \cos \theta_j,
\end{equation}

\begin{equation}
\label{f3}
      f_3=3\lambda_i \sin^4 \theta_i \cos \theta_i -3\lambda_j \sin^4 \theta_j \cos \theta_j,
\end{equation}
and
\begin{equation}
\label{f4}
      f_4= \left( 3\sin^2 \theta_i-2 \right) \cos \theta_i - \left(3\sin^2 \theta_j -2\right) \cos^2 \theta_j.
\end{equation}
For end-middle contacts, $f_1$, $f_2$, $f_3$, and $f_4$ obey different expressions.  We find
\begin{equation}
\label{f1mid}
      f_1=y_{ij}-\lambda_i \sin^3 \theta_i +\lambda_j \sin^3 \theta_j +f_5 \cos \theta_j,
\end{equation}

\begin{equation}
\label{f2mid}
      f_2=\lambda_i \sin^2 \theta_i \cos \theta_i-\lambda_j \sin^2 \theta_j \cos \theta_j + f_5 \sin \theta_j,
\end{equation}
\begin{eqnarray}
\label{f3mid}
      f_3 & = &3\lambda_i \sin^4 \theta_i \cos \theta_i-3\lambda_j \sin^4 \theta_j \cos \theta_j \nonumber\\ 
      & &  + 2f_5 \sin^3 \theta_j + f_6 \cos \theta_j,
\end{eqnarray}
and
\begin{eqnarray}
\label{f4mid}
      f_4 & = & \left(3\sin^2 \theta_i -2\right) \cos \theta_i-\left(3\sin^2 \theta_j-2 \right) \cos^2 \theta_j \nonumber \\
      & & - 2f_5 \sin^2 \theta_j \cos \theta_j + f_6 \sin \theta_j,
\end{eqnarray}
where 
\begin{eqnarray}
\label{f5mid}
      f_5 & = & \left( y_{ij}+\lambda_j \sin^3 \theta_j \right) \cos \theta_i \nonumber \\
      & & +  \left(x_{ij}+\lambda_j \cos \theta_j \right) \sin^3 \theta_i \nonumber \\
      & & -\lambda_j \sin^2 \theta_j \cos \theta_j/\sin \theta_i \nonumber \\
      & & -\left(y_{ij}+\lambda_j \sin \theta_j \right)\sin^2 \theta_i \cos \theta_i
\end{eqnarray}
and
\begin{eqnarray}
\label{f6mid}
      f_6 & = &-3\lambda_j \cos \theta_i \sin^4 \theta_j \cos \theta_j \nonumber \\
      & & +2\left( y_{ij}+\lambda_j \sin^3 \theta_j \right) \sin^3 \theta_i \nonumber \\
      & & -3\left( x_{ij}+\lambda_j \cos \theta_j \right) \sin^4 \theta_i \cos \theta_i \nonumber \\
      & & -\lambda_j \sin^3  \theta_j \left( 3\sin^2 \theta_j-2 \right) \sin \theta_i \nonumber \\
      & & +2\lambda_j \sin^2 \theta_j \cos \theta_j \sin^2 \theta_i \cos \theta_i \nonumber \\
      & & -\left(y_{ij}+\lambda_j \cos \theta_j \right)\sin^3 \theta_i \left(3\sin^2 \theta_i-2\right).
\end{eqnarray}

\nocite{*}

\bibliography{ref}

\begin{thebibliography}{38}%
\makeatletter
\providecommand \@ifxundefined [1]{%
 \@ifx{#1\undefined}
}%
\providecommand \@ifnum [1]{%
 \ifnum #1\expandafter \@firstoftwo
 \else \expandafter \@secondoftwo
 \fi
}%
\providecommand \@ifx [1]{%
 \ifx #1\expandafter \@firstoftwo
 \else \expandafter \@secondoftwo
 \fi
}%
\providecommand \natexlab [1]{#1}%
\providecommand \enquote  [1]{``#1''}%
\providecommand \bibnamefont  [1]{#1}%
\providecommand \bibfnamefont [1]{#1}%
\providecommand \citenamefont [1]{#1}%
\providecommand \href@noop [0]{\@secondoftwo}%
\providecommand \href [0]{\begingroup \@sanitize@url \@href}%
\providecommand \@href[1]{\@@startlink{#1}\@@href}%
\providecommand \@@href[1]{\endgroup#1\@@endlink}%
\providecommand \@sanitize@url [0]{\catcode `\\12\catcode `\$12\catcode
  `\&12\catcode `\#12\catcode `\^12\catcode `\_12\catcode `\%12\relax}%
\providecommand \@@startlink[1]{}%
\providecommand \@@endlink[0]{}%
\providecommand \url  [0]{\begingroup\@sanitize@url \@url }%
\providecommand \@url [1]{\endgroup\@href {#1}{\urlprefix }}%
\providecommand \urlprefix  [0]{URL }%
\providecommand \Eprint [0]{\href }%
\providecommand \doibase [0]{https://doi.org/}%
\providecommand \selectlanguage [0]{\@gobble}%
\providecommand \bibinfo  [0]{\@secondoftwo}%
\providecommand \bibfield  [0]{\@secondoftwo}%
\providecommand \translation [1]{[#1]}%
\providecommand \BibitemOpen [0]{}%
\providecommand \bibitemStop [0]{}%
\providecommand \bibitemNoStop [0]{.\EOS\space}%
\providecommand \EOS [0]{\spacefactor3000\relax}%
\providecommand \BibitemShut  [1]{\csname bibitem#1\endcsname}%
\let\auto@bib@innerbib\@empty
\bibitem [{\citenamefont {Jaeger}\ \emph {et~al.}(1996)\citenamefont {Jaeger},
  \citenamefont {Nagel},\ and\ \citenamefont {Behringer}}]{rmp}%
  \BibitemOpen
  \bibfield  {author} {\bibinfo {author} {\bibfnamefont {H.~M.}\ \bibnamefont
  {Jaeger}}, \bibinfo {author} {\bibfnamefont {S.~R.}\ \bibnamefont {Nagel}},\
  and\ \bibinfo {author} {\bibfnamefont {R.~P.}\ \bibnamefont {Behringer}},\
  }\bibfield  {title} {\bibinfo {title} {Granular solids, liquids, and gases},\
  }\href@noop {} {\bibfield  {journal} {\bibinfo  {journal} {Rev. Mod. Phys.}\
  }\textbf {\bibinfo {volume} {68}},\ \bibinfo {pages} {1259} (\bibinfo {year}
  {1996})}\BibitemShut {NoStop}%
\bibitem [{\citenamefont {Behringer}\ and\ \citenamefont
  {Chakraborty}(2018)}]{phys_jamming}%
  \BibitemOpen
  \bibfield  {author} {\bibinfo {author} {\bibfnamefont {R.~P.}\ \bibnamefont
  {Behringer}}\ and\ \bibinfo {author} {\bibfnamefont {B.}~\bibnamefont
  {Chakraborty}},\ }\bibfield  {title} {\bibinfo {title} {The physics of
  jamming for granular materials: A review},\ }\href@noop {} {\bibfield
  {journal} {\bibinfo  {journal} {Rep. Prog. Phys.}\ }\textbf {\bibinfo
  {volume} {82}},\ \bibinfo {pages} {012601} (\bibinfo {year}
  {2018})}\BibitemShut {NoStop}%
\bibitem [{\citenamefont {Cain}\ \emph {et~al.}(2001)\citenamefont {Cain},
  \citenamefont {Page},\ and\ \citenamefont {Biggs}}]{cain}%
  \BibitemOpen
  \bibfield  {author} {\bibinfo {author} {\bibfnamefont {R.~G.}\ \bibnamefont
  {Cain}}, \bibinfo {author} {\bibfnamefont {N.~W.}\ \bibnamefont {Page}},\
  and\ \bibinfo {author} {\bibfnamefont {S.}~\bibnamefont {Biggs}},\ }\bibfield
   {title} {\bibinfo {title} {Microscopic and macroscopic aspects of stick-slip
  motion in granular shear},\ }\href@noop {} {\bibfield  {journal} {\bibinfo
  {journal} {Phys. Rev. E}\ }\textbf {\bibinfo {volume} {64}},\ \bibinfo
  {pages} {016413} (\bibinfo {year} {2001})}\BibitemShut {NoStop}%
\bibitem [{\citenamefont {Fenistein}\ and\ \citenamefont {van
  Hecke}(2003)}]{fenistein}%
  \BibitemOpen
  \bibfield  {author} {\bibinfo {author} {\bibfnamefont {D.}~\bibnamefont
  {Fenistein}}\ and\ \bibinfo {author} {\bibfnamefont {M.}~\bibnamefont {van
  Hecke}},\ }\bibfield  {title} {\bibinfo {title} {Wide shear zones in granular
  bulk flow},\ }\href@noop {} {\bibfield  {journal} {\bibinfo  {journal}
  {Nature}\ }\textbf {\bibinfo {volume} {425}},\ \bibinfo {pages} {256}
  (\bibinfo {year} {2003})}\BibitemShut {NoStop}%
\bibitem [{\citenamefont {Hill}\ \emph {et~al.}(1997)\citenamefont {Hill},
  \citenamefont {Caprihan},\ and\ \citenamefont {Kakalios}}]{segregation}%
  \BibitemOpen
  \bibfield  {author} {\bibinfo {author} {\bibfnamefont {K.~M.}\ \bibnamefont
  {Hill}}, \bibinfo {author} {\bibfnamefont {A.}~\bibnamefont {Caprihan}},\
  and\ \bibinfo {author} {\bibfnamefont {J.}~\bibnamefont {Kakalios}},\
  }\bibfield  {title} {\bibinfo {title} {Axial segregation of granular media
  rotated in a drum mixer: Pattern evolution},\ }\href@noop {} {\bibfield
  {journal} {\bibinfo  {journal} {Phys. Rev. E}\ }\textbf {\bibinfo {volume}
  {56}},\ \bibinfo {pages} {4386} (\bibinfo {year} {1997})}\BibitemShut
  {NoStop}%
\bibitem [{\citenamefont {Ben-Zion}\ and\ \citenamefont
  {Sammis}(2011)}]{benzion}%
  \BibitemOpen
  \bibfield  {author} {\bibinfo {author} {\bibfnamefont {Y.}~\bibnamefont
  {Ben-Zion}}\ and\ \bibinfo {author} {\bibfnamefont {C.}~\bibnamefont
  {Sammis}},\ }\bibfield  {title} {\bibinfo {title} {Brittle deformation of
  solid and granular materials with applications to mechanics of earthquakes
  and faults},\ }\href@noop {} {\bibfield  {journal} {\bibinfo  {journal} {Pure
  and Applied Geophysics}\ }\textbf {\bibinfo {volume} {168}},\ \bibinfo
  {pages} {2147} (\bibinfo {year} {2011})}\BibitemShut {NoStop}%
\bibitem [{\citenamefont {Dahmen}\ \emph {et~al.}(2011)\citenamefont {Dahmen},
  \citenamefont {Ben-Zion},\ and\ \citenamefont {Uhl}}]{benzion2}%
  \BibitemOpen
  \bibfield  {author} {\bibinfo {author} {\bibfnamefont {K.~A.}\ \bibnamefont
  {Dahmen}}, \bibinfo {author} {\bibfnamefont {Y.}~\bibnamefont {Ben-Zion}},\
  and\ \bibinfo {author} {\bibfnamefont {J.~T.}\ \bibnamefont {Uhl}},\
  }\bibfield  {title} {\bibinfo {title} {A simple analytic theory for the
  statistics of avalanches in sheared granular materials},\ }\href@noop {}
  {\bibfield  {journal} {\bibinfo  {journal} {Nature Physics}\ }\textbf
  {\bibinfo {volume} {7}},\ \bibinfo {pages} {554} (\bibinfo {year}
  {2011})}\BibitemShut {NoStop}%
\bibitem [{\citenamefont {Chen}\ \emph {et~al.}(2021)\citenamefont {Chen},
  \citenamefont {Wassgren},\ and\ \citenamefont {Ambrose}}]{chen}%
  \BibitemOpen
  \bibfield  {author} {\bibinfo {author} {\bibfnamefont {Z.}~\bibnamefont
  {Chen}}, \bibinfo {author} {\bibfnamefont {C.}~\bibnamefont {Wassgren}},\
  and\ \bibinfo {author} {\bibfnamefont {R.~P.~K.}\ \bibnamefont {Ambrose}},\
  }\bibfield  {title} {\bibinfo {title} {Measured damage resistance of corn and
  wheat kernels to compression, friction, repeated impacts},\ }\href@noop {}
  {\bibfield  {journal} {\bibinfo  {journal} {Powder Technology}\ }\textbf
  {\bibinfo {volume} {380}},\ \bibinfo {pages} {638} (\bibinfo {year}
  {2021})}\BibitemShut {NoStop}%
\bibitem [{\citenamefont {Muzzio}\ \emph {et~al.}(2003)\citenamefont {Muzzio},
  \citenamefont {Goodridge}, \citenamefont {Alexander}, \citenamefont
  {Arratia}, \citenamefont {Yang}, \citenamefont {Sudah},\ and\ \citenamefont
  {Mergen}}]{muzzio}%
  \BibitemOpen
  \bibfield  {author} {\bibinfo {author} {\bibfnamefont {F.~J.}\ \bibnamefont
  {Muzzio}}, \bibinfo {author} {\bibfnamefont {C.~L.}\ \bibnamefont
  {Goodridge}}, \bibinfo {author} {\bibfnamefont {A.}~\bibnamefont
  {Alexander}}, \bibinfo {author} {\bibfnamefont {P.}~\bibnamefont {Arratia}},
  \bibinfo {author} {\bibfnamefont {H.}~\bibnamefont {Yang}}, \bibinfo {author}
  {\bibfnamefont {O.}~\bibnamefont {Sudah}},\ and\ \bibinfo {author}
  {\bibfnamefont {G.}~\bibnamefont {Mergen}},\ }\bibfield  {title} {\bibinfo
  {title} {Sampling and characterization of pharmaceutical powders and granular
  beds},\ }\href@noop {} {\bibfield  {journal} {\bibinfo  {journal}
  {International Journal of Pharmaceutics}\ }\textbf {\bibinfo {volume}
  {250}},\ \bibinfo {pages} {51} (\bibinfo {year} {2003})}\BibitemShut
  {NoStop}%
\bibitem [{\citenamefont {Brown}\ \emph {et~al.}(2010)\citenamefont {Brown},
  \citenamefont {Rodenberg}, \citenamefont {Amend}, \citenamefont {Mozeika},
  \citenamefont {Steltz}, \citenamefont {Zakin}, \citenamefont {Lipson},\ and\
  \citenamefont {Jaeger}}]{brown}%
  \BibitemOpen
  \bibfield  {author} {\bibinfo {author} {\bibfnamefont {E.}~\bibnamefont
  {Brown}}, \bibinfo {author} {\bibfnamefont {N.}~\bibnamefont {Rodenberg}},
  \bibinfo {author} {\bibfnamefont {J.}~\bibnamefont {Amend}}, \bibinfo
  {author} {\bibfnamefont {A.}~\bibnamefont {Mozeika}}, \bibinfo {author}
  {\bibfnamefont {E.}~\bibnamefont {Steltz}}, \bibinfo {author} {\bibfnamefont
  {M.~R.}\ \bibnamefont {Zakin}}, \bibinfo {author} {\bibfnamefont
  {H.}~\bibnamefont {Lipson}},\ and\ \bibinfo {author} {\bibfnamefont {H.~M.}\
  \bibnamefont {Jaeger}},\ }\bibfield  {title} {\bibinfo {title} {Universal
  robotic gripper based on the jamming of granular material},\ }\href@noop {}
  {\bibfield  {journal} {\bibinfo  {journal} {Proceedings of the National
  Academy of Sciences, USA}\ }\textbf {\bibinfo {volume} {107}},\ \bibinfo
  {pages} {18809} (\bibinfo {year} {2010})}\BibitemShut {NoStop}%
\bibitem [{\citenamefont {Shah}\ \emph {et~al.}(2020)\citenamefont {Shah},
  \citenamefont {Yang}, \citenamefont {Yuen}, \citenamefont {Huang},\ and\
  \citenamefont {Kramer-Bottiglio}}]{kramer}%
  \BibitemOpen
  \bibfield  {author} {\bibinfo {author} {\bibfnamefont {D.~S.}\ \bibnamefont
  {Shah}}, \bibinfo {author} {\bibfnamefont {E.~J.}\ \bibnamefont {Yang}},
  \bibinfo {author} {\bibfnamefont {M.~C.}\ \bibnamefont {Yuen}}, \bibinfo
  {author} {\bibfnamefont {E.~C.}\ \bibnamefont {Huang}},\ and\ \bibinfo
  {author} {\bibfnamefont {R.}~\bibnamefont {Kramer-Bottiglio}},\ }\bibfield
  {title} {\bibinfo {title} {Jamming skins that control system rigidity from
  the surface},\ }\href@noop {} {\bibfield  {journal} {\bibinfo  {journal}
  {Adv. Funct. Mater.}\ ,\ \bibinfo {pages} {2006915}} (\bibinfo {year}
  {2020})}\BibitemShut {NoStop}%
\bibitem [{\citenamefont {O'Hern}\ \emph {et~al.}(2003)\citenamefont {O'Hern},
  \citenamefont {Silbert}, \citenamefont {Liu},\ and\ \citenamefont
  {Nagel}}]{epitome}%
  \BibitemOpen
  \bibfield  {author} {\bibinfo {author} {\bibfnamefont {C.~S.}\ \bibnamefont
  {O'Hern}}, \bibinfo {author} {\bibfnamefont {L.~E.}\ \bibnamefont {Silbert}},
  \bibinfo {author} {\bibfnamefont {A.~J.}\ \bibnamefont {Liu}},\ and\ \bibinfo
  {author} {\bibfnamefont {S.~R.}\ \bibnamefont {Nagel}},\ }\bibfield  {title}
  {\bibinfo {title} {Jamming at zero temperature and zero applied stress: The
  epitome of disorder},\ }\href@noop {} {\bibfield  {journal} {\bibinfo
  {journal} {Phys. Rev. E}\ }\textbf {\bibinfo {volume} {68}},\ \bibinfo
  {pages} {011306} (\bibinfo {year} {2003})}\BibitemShut {NoStop}%
\bibitem [{\citenamefont {Durian}(1995)}]{dd}%
  \BibitemOpen
  \bibfield  {author} {\bibinfo {author} {\bibfnamefont {D.~J.}\ \bibnamefont
  {Durian}},\ }\bibfield  {title} {\bibinfo {title} {Foam mechanics at the
  bubble scale},\ }\href@noop {} {\bibfield  {journal} {\bibinfo  {journal}
  {Phys. Rev. Lett.}\ }\textbf {\bibinfo {volume} {75}},\ \bibinfo {pages}
  {4780} (\bibinfo {year} {1995})}\BibitemShut {NoStop}%
\bibitem [{\citenamefont {Schreck}\ \emph {et~al.}(2014)\citenamefont
  {Schreck}, \citenamefont {O'Hern},\ and\ \citenamefont {Shattuck}}]{gm}%
  \BibitemOpen
  \bibfield  {author} {\bibinfo {author} {\bibfnamefont {C.~F.}\ \bibnamefont
  {Schreck}}, \bibinfo {author} {\bibfnamefont {C.~S.}\ \bibnamefont
  {O'Hern}},\ and\ \bibinfo {author} {\bibfnamefont {M.~D.}\ \bibnamefont
  {Shattuck}},\ }\bibfield  {title} {\bibinfo {title} {Vibrations of jammed
  disk packings with {H}ertzian interacations},\ }\href@noop {} {\bibfield
  {journal} {\bibinfo  {journal} {Granular Matter}\ }\textbf {\bibinfo {volume}
  {16}},\ \bibinfo {pages} {209} (\bibinfo {year} {2014})}\BibitemShut
  {NoStop}%
\bibitem [{\citenamefont {Tkachenko}\ and\ \citenamefont
  {Witten}(1999)}]{witten}%
  \BibitemOpen
  \bibfield  {author} {\bibinfo {author} {\bibfnamefont {A.~V.}\ \bibnamefont
  {Tkachenko}}\ and\ \bibinfo {author} {\bibfnamefont {T.~A.}\ \bibnamefont
  {Witten}},\ }\bibfield  {title} {\bibinfo {title} {Stress propagation through
  frictionless granular material},\ }\href@noop {} {\bibfield  {journal}
  {\bibinfo  {journal} {Phys. Rev. E}\ }\textbf {\bibinfo {volume} {60}},\
  \bibinfo {pages} {687} (\bibinfo {year} {1999})}\BibitemShut {NoStop}%
\bibitem [{\citenamefont {Makse}\ \emph {et~al.}(2000)\citenamefont {Makse},
  \citenamefont {Johnson},\ and\ \citenamefont {Schwartz}}]{makse}%
  \BibitemOpen
  \bibfield  {author} {\bibinfo {author} {\bibfnamefont {H.~A.}\ \bibnamefont
  {Makse}}, \bibinfo {author} {\bibfnamefont {D.~L.}\ \bibnamefont {Johnson}},\
  and\ \bibinfo {author} {\bibfnamefont {L.~M.}\ \bibnamefont {Schwartz}},\
  }\bibfield  {title} {\bibinfo {title} {Packing of compressible granular
  materials},\ }\href@noop {} {\bibfield  {journal} {\bibinfo  {journal} {Phys.
  Rev. Lett.}\ }\textbf {\bibinfo {volume} {84}},\ \bibinfo {pages} {4160}
  (\bibinfo {year} {2000})}\BibitemShut {NoStop}%
\bibitem [{\citenamefont {Goodrich}\ \emph {et~al.}(2012)\citenamefont
  {Goodrich}, \citenamefont {Liu},\ and\ \citenamefont {Nagel}}]{goodrich}%
  \BibitemOpen
  \bibfield  {author} {\bibinfo {author} {\bibfnamefont {C.~P.}\ \bibnamefont
  {Goodrich}}, \bibinfo {author} {\bibfnamefont {A.~J.}\ \bibnamefont {Liu}},\
  and\ \bibinfo {author} {\bibfnamefont {S.~R.}\ \bibnamefont {Nagel}},\
  }\bibfield  {title} {\bibinfo {title} {Finite-size scaling at the jamming
  transition},\ }\href@noop {} {\bibfield  {journal} {\bibinfo  {journal}
  {Phys. Rev. Lett.}\ }\textbf {\bibinfo {volume} {109}},\ \bibinfo {pages}
  {095704} (\bibinfo {year} {2012})}\BibitemShut {NoStop}%
\bibitem [{\citenamefont {Mason}\ \emph {et~al.}(1995)\citenamefont {Mason},
  \citenamefont {Bibette},\ and\ \citenamefont {Weitz}}]{mason}%
  \BibitemOpen
  \bibfield  {author} {\bibinfo {author} {\bibfnamefont {T.~G.}\ \bibnamefont
  {Mason}}, \bibinfo {author} {\bibfnamefont {J.}~\bibnamefont {Bibette}},\
  and\ \bibinfo {author} {\bibfnamefont {D.~A.}\ \bibnamefont {Weitz}},\
  }\bibfield  {title} {\bibinfo {title} {Elasticity of compressed emulsions},\
  }\href@noop {} {\bibfield  {journal} {\bibinfo  {journal} {Phys. Rev. Lett.}\
  }\textbf {\bibinfo {volume} {75}},\ \bibinfo {pages} {2051} (\bibinfo {year}
  {1995})}\BibitemShut {NoStop}%
\bibitem [{\citenamefont {Jorjadze}\ \emph {et~al.}(2013)\citenamefont
  {Jorjadze}, \citenamefont {Pontani},\ and\ \citenamefont {Brujic}}]{brujic}%
  \BibitemOpen
  \bibfield  {author} {\bibinfo {author} {\bibfnamefont {I.}~\bibnamefont
  {Jorjadze}}, \bibinfo {author} {\bibfnamefont {L.-L.}\ \bibnamefont
  {Pontani}},\ and\ \bibinfo {author} {\bibfnamefont {J.}~\bibnamefont
  {Brujic}},\ }\bibfield  {title} {\bibinfo {title} {Microscopic approach to
  the nonlinear elasticity of compressed emulsions},\ }\href@noop {} {\bibfield
   {journal} {\bibinfo  {journal} {Phys. Rev. Lett.}\ }\textbf {\bibinfo
  {volume} {110}},\ \bibinfo {pages} {048302} (\bibinfo {year}
  {2013})}\BibitemShut {NoStop}%
\bibitem [{\citenamefont {Majmudar}\ \emph {et~al.}(2007)\citenamefont
  {Majmudar}, \citenamefont {Sperl}, \citenamefont {Luding},\ and\
  \citenamefont {Behringer}}]{behringer}%
  \BibitemOpen
  \bibfield  {author} {\bibinfo {author} {\bibfnamefont {T.~S.}\ \bibnamefont
  {Majmudar}}, \bibinfo {author} {\bibfnamefont {M.}~\bibnamefont {Sperl}},
  \bibinfo {author} {\bibfnamefont {S.}~\bibnamefont {Luding}},\ and\ \bibinfo
  {author} {\bibfnamefont {R.~P.}\ \bibnamefont {Behringer}},\ }\bibfield
  {title} {\bibinfo {title} {Jamming transition in granular systems},\
  }\href@noop {} {\bibfield  {journal} {\bibinfo  {journal} {Phys. Rev. Lett.}\
  }\textbf {\bibinfo {volume} {98}},\ \bibinfo {pages} {058001} (\bibinfo
  {year} {2007})}\BibitemShut {NoStop}%
\bibitem [{\citenamefont {VanderWerf}\ \emph {et~al.}(2020)\citenamefont
  {VanderWerf}, \citenamefont {Boromand}, \citenamefont {Shattuck},\ and\
  \citenamefont {O'Hern}}]{Kyle_PRL}%
  \BibitemOpen
  \bibfield  {author} {\bibinfo {author} {\bibfnamefont {K.}~\bibnamefont
  {VanderWerf}}, \bibinfo {author} {\bibfnamefont {A.}~\bibnamefont
  {Boromand}}, \bibinfo {author} {\bibfnamefont {M.~D.}\ \bibnamefont
  {Shattuck}},\ and\ \bibinfo {author} {\bibfnamefont {C.~S.}\ \bibnamefont
  {O'Hern}},\ }\bibfield  {title} {\bibinfo {title} {Pressure-dependent shear
  response of jammed packings of spherical particles},\ }\href@noop {}
  {\bibfield  {journal} {\bibinfo  {journal} {Phys. Rev. Lett.}\ }\textbf
  {\bibinfo {volume} {124}},\ \bibinfo {pages} {038004} (\bibinfo {year}
  {2020})}\BibitemShut {NoStop}%
\bibitem [{\citenamefont {Tuckman}\ \emph {et~al.}(2020)\citenamefont
  {Tuckman}, \citenamefont {VanderWerf}, \citenamefont {Yuan}, \citenamefont
  {Zhang}, \citenamefont {Zhang}, \citenamefont {Shattuck},\ and\ \citenamefont
  {O'Hern}}]{PJ_jump_changes}%
  \BibitemOpen
  \bibfield  {author} {\bibinfo {author} {\bibfnamefont {P.~J.}\ \bibnamefont
  {Tuckman}}, \bibinfo {author} {\bibfnamefont {K.}~\bibnamefont {VanderWerf}},
  \bibinfo {author} {\bibfnamefont {Y.}~\bibnamefont {Yuan}}, \bibinfo {author}
  {\bibfnamefont {S.}~\bibnamefont {Zhang}}, \bibinfo {author} {\bibfnamefont
  {J.}~\bibnamefont {Zhang}}, \bibinfo {author} {\bibfnamefont {M.~D.}\
  \bibnamefont {Shattuck}},\ and\ \bibinfo {author} {\bibfnamefont {C.~S.}\
  \bibnamefont {O'Hern}},\ }\bibfield  {title} {\bibinfo {title} {Contact
  network changes in ordered and disordered disk packings},\ }\href@noop {}
  {\bibfield  {journal} {\bibinfo  {journal} {Soft Matter}\ }\textbf {\bibinfo
  {volume} {16}},\ \bibinfo {pages} {9443} (\bibinfo {year}
  {2020})}\BibitemShut {NoStop}%
\bibitem [{\citenamefont {Morse}\ \emph {et~al.}(2020)\citenamefont {Morse},
  \citenamefont {Wijtmans}, \citenamefont {van Deen}, \citenamefont {van
  Hecke},\ and\ \citenamefont {Manning}}]{plasticity}%
  \BibitemOpen
  \bibfield  {author} {\bibinfo {author} {\bibfnamefont {P.}~\bibnamefont
  {Morse}}, \bibinfo {author} {\bibfnamefont {S.}~\bibnamefont {Wijtmans}},
  \bibinfo {author} {\bibfnamefont {M.}~\bibnamefont {van Deen}}, \bibinfo
  {author} {\bibfnamefont {M.}~\bibnamefont {van Hecke}},\ and\ \bibinfo
  {author} {\bibfnamefont {M.~L.}\ \bibnamefont {Manning}},\ }\bibfield
  {title} {\bibinfo {title} {Differences in plasticity between hard and soft
  spheres},\ }\href {https://doi.org/10.1103/PhysRevResearch.2.023179}
  {\bibfield  {journal} {\bibinfo  {journal} {Phys. Rev. Research}\ }\textbf
  {\bibinfo {volume} {2}},\ \bibinfo {pages} {023179} (\bibinfo {year}
  {2020})}\BibitemShut {NoStop}%
\bibitem [{\citenamefont {Wang}\ \emph {et~al.}(2021)\citenamefont {Wang},
  \citenamefont {Zhang}, \citenamefont {Tuckman}, \citenamefont {Ouellette},
  \citenamefont {Shattuck},\ and\ \citenamefont {O'Hern}}]{Philip_Herzian}%
  \BibitemOpen
  \bibfield  {author} {\bibinfo {author} {\bibfnamefont {P.}~\bibnamefont
  {Wang}}, \bibinfo {author} {\bibfnamefont {S.}~\bibnamefont {Zhang}},
  \bibinfo {author} {\bibfnamefont {P.~J.}\ \bibnamefont {Tuckman}}, \bibinfo
  {author} {\bibfnamefont {N.~T.}\ \bibnamefont {Ouellette}}, \bibinfo {author}
  {\bibfnamefont {M.~D.}\ \bibnamefont {Shattuck}},\ and\ \bibinfo {author}
  {\bibfnamefont {C.~S.}\ \bibnamefont {O'Hern}},\ }\bibfield  {title}
  {\bibinfo {title} {Shear response of granular packings compressed above
  jamming onset},\ }\href@noop {} {\bibfield  {journal} {\bibinfo  {journal}
  {Phys. Rev. E}\ }\textbf {\bibinfo {volume} {103}},\ \bibinfo {pages}
  {022902} (\bibinfo {year} {2021})}\BibitemShut {NoStop}%
\bibitem [{\citenamefont {Mailman}\ \emph {et~al.}(2009)\citenamefont
  {Mailman}, \citenamefont {Schreck}, \citenamefont {O'Hern},\ and\
  \citenamefont {Chakraborty}}]{mailman}%
  \BibitemOpen
  \bibfield  {author} {\bibinfo {author} {\bibfnamefont {M.}~\bibnamefont
  {Mailman}}, \bibinfo {author} {\bibfnamefont {C.~F.}\ \bibnamefont
  {Schreck}}, \bibinfo {author} {\bibfnamefont {C.~S.}\ \bibnamefont
  {O'Hern}},\ and\ \bibinfo {author} {\bibfnamefont {B.}~\bibnamefont
  {Chakraborty}},\ }\bibfield  {title} {\bibinfo {title} {Jamming in systems
  composed of frictionless ellipse-shaped particles},\ }\href@noop {}
  {\bibfield  {journal} {\bibinfo  {journal} {Phys. Rev. Lett.}\ }\textbf
  {\bibinfo {volume} {102}},\ \bibinfo {pages} {255501} (\bibinfo {year}
  {2009})}\BibitemShut {NoStop}%
\bibitem [{\citenamefont {Schreck}\ \emph {et~al.}(2010)\citenamefont
  {Schreck}, \citenamefont {Xu},\ and\ \citenamefont
  {O'Hern}}]{Shreck_dimer_ellipse}%
  \BibitemOpen
  \bibfield  {author} {\bibinfo {author} {\bibfnamefont {C.~F.}\ \bibnamefont
  {Schreck}}, \bibinfo {author} {\bibfnamefont {N.}~\bibnamefont {Xu}},\ and\
  \bibinfo {author} {\bibfnamefont {C.~S.}\ \bibnamefont {O'Hern}},\ }\bibfield
   {title} {\bibinfo {title} {A comparison of jamming behavior in systems
  composed of dimer- and ellipse-shaped particles},\ }\href@noop {} {\bibfield
  {journal} {\bibinfo  {journal} {Soft Matter}\ }\textbf {\bibinfo {volume}
  {6}},\ \bibinfo {pages} {2960} (\bibinfo {year} {2010})}\BibitemShut
  {NoStop}%
\bibitem [{\citenamefont {VanderWerf}\ \emph {et~al.}(2018)\citenamefont
  {VanderWerf}, \citenamefont {Jin}, \citenamefont {Shattuck},\ and\
  \citenamefont {O'Hern}}]{Kyle_nonsphere}%
  \BibitemOpen
  \bibfield  {author} {\bibinfo {author} {\bibfnamefont {K.}~\bibnamefont
  {VanderWerf}}, \bibinfo {author} {\bibfnamefont {W.}~\bibnamefont {Jin}},
  \bibinfo {author} {\bibfnamefont {M.~D.}\ \bibnamefont {Shattuck}},\ and\
  \bibinfo {author} {\bibfnamefont {C.~S.}\ \bibnamefont {O'Hern}},\ }\bibfield
   {title} {\bibinfo {title} {Hypostatic jammed packings of frictionless
  nonspherical particles},\ }\href@noop {} {\bibfield  {journal} {\bibinfo
  {journal} {Phys. Rev. E}\ }\textbf {\bibinfo {volume} {97}},\ \bibinfo
  {pages} {012909} (\bibinfo {year} {2018})}\BibitemShut {NoStop}%
\bibitem [{\citenamefont {Chen}\ \emph {et~al.}(2018)\citenamefont {Chen},
  \citenamefont {Bertrand}, \citenamefont {Jin}, \citenamefont {Shattuck},\
  and\ \citenamefont {O'Hern}}]{Sheng_stress_anisotropy}%
  \BibitemOpen
  \bibfield  {author} {\bibinfo {author} {\bibfnamefont {S.}~\bibnamefont
  {Chen}}, \bibinfo {author} {\bibfnamefont {T.}~\bibnamefont {Bertrand}},
  \bibinfo {author} {\bibfnamefont {W.}~\bibnamefont {Jin}}, \bibinfo {author}
  {\bibfnamefont {M.~D.}\ \bibnamefont {Shattuck}},\ and\ \bibinfo {author}
  {\bibfnamefont {C.~S.}\ \bibnamefont {O'Hern}},\ }\bibfield  {title}
  {\bibinfo {title} {Stress anisotropy in shear-jammed packings of frictionless
  disks},\ }\href@noop {} {\bibfield  {journal} {\bibinfo  {journal} {Phys.
  Rev. E}\ }\textbf {\bibinfo {volume} {98}},\ \bibinfo {pages} {042906}
  (\bibinfo {year} {2018})}\BibitemShut {NoStop}%
\bibitem [{\citenamefont {Edwards}\ and\ \citenamefont
  {Grinev}(2001)}]{Love_expression}%
  \BibitemOpen
  \bibfield  {author} {\bibinfo {author} {\bibfnamefont {S.~F.}\ \bibnamefont
  {Edwards}}\ and\ \bibinfo {author} {\bibfnamefont {D.~V.}\ \bibnamefont
  {Grinev}},\ }\bibfield  {title} {\bibinfo {title} {Transmission of stress in
  granular materials as a problem of statistical mechanics},\ }\href@noop {}
  {\bibfield  {journal} {\bibinfo  {journal} {Physica A: Statistical Mechanics
  and its Applications}\ }\textbf {\bibinfo {volume} {302}},\ \bibinfo {pages}
  {162} (\bibinfo {year} {2001})}\BibitemShut {NoStop}%
\bibitem [{\citenamefont {Bertrand}\ \emph {et~al.}(2016)\citenamefont
  {Bertrand}, \citenamefont {Behringer}, \citenamefont {Chakraborty},
  \citenamefont {O'Hern},\ and\ \citenamefont {Shattuck}}]{protocol}%
  \BibitemOpen
  \bibfield  {author} {\bibinfo {author} {\bibfnamefont {T.}~\bibnamefont
  {Bertrand}}, \bibinfo {author} {\bibfnamefont {R.~P.}\ \bibnamefont
  {Behringer}}, \bibinfo {author} {\bibfnamefont {B.}~\bibnamefont
  {Chakraborty}}, \bibinfo {author} {\bibfnamefont {C.~S.}\ \bibnamefont
  {O'Hern}},\ and\ \bibinfo {author} {\bibfnamefont {M.~D.}\ \bibnamefont
  {Shattuck}},\ }\bibfield  {title} {\bibinfo {title} {Protocol dependence of
  the jamming transition},\ }\href@noop {} {\bibfield  {journal} {\bibinfo
  {journal} {Phys. Rev. E}\ }\textbf {\bibinfo {volume} {93}},\ \bibinfo
  {pages} {012901} (\bibinfo {year} {2016})}\BibitemShut {NoStop}%
\bibitem [{\citenamefont {Mizuno}\ \emph {et~al.}(2016)\citenamefont {Mizuno},
  \citenamefont {Saitoh},\ and\ \citenamefont {Silbert}}]{mizuno2016elastic}%
  \BibitemOpen
  \bibfield  {author} {\bibinfo {author} {\bibfnamefont {H.}~\bibnamefont
  {Mizuno}}, \bibinfo {author} {\bibfnamefont {K.}~\bibnamefont {Saitoh}},\
  and\ \bibinfo {author} {\bibfnamefont {L.~E.}\ \bibnamefont {Silbert}},\
  }\bibfield  {title} {\bibinfo {title} {Elastic moduli and vibrational modes
  in jammed particulate packings},\ }\href@noop {} {\bibfield  {journal}
  {\bibinfo  {journal} {Phys. Rev. E}\ }\textbf {\bibinfo {volume} {93}},\
  \bibinfo {pages} {062905} (\bibinfo {year} {2016})}\BibitemShut {NoStop}%
\bibitem [{\citenamefont {Ellenbroek}\ \emph {et~al.}(2009)\citenamefont
  {Ellenbroek}, \citenamefont {Zeravcic}, \citenamefont {van Saarloos},\ and\
  \citenamefont {van Hecke}}]{ellenbroek2009non}%
  \BibitemOpen
  \bibfield  {author} {\bibinfo {author} {\bibfnamefont {W.~G.}\ \bibnamefont
  {Ellenbroek}}, \bibinfo {author} {\bibfnamefont {Z.}~\bibnamefont
  {Zeravcic}}, \bibinfo {author} {\bibfnamefont {W.}~\bibnamefont {van
  Saarloos}},\ and\ \bibinfo {author} {\bibfnamefont {M.}~\bibnamefont {van
  Hecke}},\ }\bibfield  {title} {\bibinfo {title} {Non-affine response: Jammed
  packings vs. spring networks},\ }\href@noop {} {\bibfield  {journal}
  {\bibinfo  {journal} {EPL}\ }\textbf {\bibinfo {volume} {87}},\ \bibinfo
  {pages} {34004} (\bibinfo {year} {2009})}\BibitemShut {NoStop}%
\bibitem [{\citenamefont {Schlegel}\ \emph {et~al.}(2016)\citenamefont
  {Schlegel}, \citenamefont {Brujic}, \citenamefont {Terentjev},\ and\
  \citenamefont {Zaccone}}]{schlegel2016local}%
  \BibitemOpen
  \bibfield  {author} {\bibinfo {author} {\bibfnamefont {M.}~\bibnamefont
  {Schlegel}}, \bibinfo {author} {\bibfnamefont {J.}~\bibnamefont {Brujic}},
  \bibinfo {author} {\bibfnamefont {E.}~\bibnamefont {Terentjev}},\ and\
  \bibinfo {author} {\bibfnamefont {A.}~\bibnamefont {Zaccone}},\ }\bibfield
  {title} {\bibinfo {title} {Local structure controls the nonaffine shear and
  bulk moduli of disordered solids},\ }\href@noop {} {\bibfield  {journal}
  {\bibinfo  {journal} {Scientific Reports}\ }\textbf {\bibinfo {volume} {6}},\
  \bibinfo {pages} {1} (\bibinfo {year} {2016})}\BibitemShut {NoStop}%
\bibitem [{\citenamefont {Zaccone}\ and\ \citenamefont
  {Scossa-Romano}(2011)}]{zaccone2011approximate}%
  \BibitemOpen
  \bibfield  {author} {\bibinfo {author} {\bibfnamefont {A.}~\bibnamefont
  {Zaccone}}\ and\ \bibinfo {author} {\bibfnamefont {E.}~\bibnamefont
  {Scossa-Romano}},\ }\bibfield  {title} {\bibinfo {title} {Approximate
  analytical description of the nonaffine response of amorphous solids},\
  }\href@noop {} {\bibfield  {journal} {\bibinfo  {journal} {Phys. Rev. B}\
  }\textbf {\bibinfo {volume} {83}},\ \bibinfo {pages} {184205} (\bibinfo
  {year} {2011})}\BibitemShut {NoStop}%
\bibitem [{\citenamefont {Phillips}\ \emph {et~al.}(2021)\citenamefont
  {Phillips}, \citenamefont {Baggioli}, \citenamefont {Sirk}, \citenamefont
  {Trachenko},\ and\ \citenamefont {Zaccone}}]{phillips2021universal}%
  \BibitemOpen
  \bibfield  {author} {\bibinfo {author} {\bibfnamefont {A.~E.}\ \bibnamefont
  {Phillips}}, \bibinfo {author} {\bibfnamefont {M.}~\bibnamefont {Baggioli}},
  \bibinfo {author} {\bibfnamefont {T.~W.}\ \bibnamefont {Sirk}}, \bibinfo
  {author} {\bibfnamefont {K.}~\bibnamefont {Trachenko}},\ and\ \bibinfo
  {author} {\bibfnamefont {A.}~\bibnamefont {Zaccone}},\ }\bibfield  {title}
  {\bibinfo {title} {Universal ${L}^{-3}$ finite-size effects in the
  viscoelasticity of amorphous systems},\ }\href@noop {} {\bibfield  {journal}
  {\bibinfo  {journal} {Phys. Rev. Mater.}\ }\textbf {\bibinfo {volume} {5}},\
  \bibinfo {pages} {035602} (\bibinfo {year} {2021})}\BibitemShut {NoStop}%
\bibitem [{\citenamefont {Papanikolaou}\ \emph {et~al.}(2013)\citenamefont
  {Papanikolaou}, \citenamefont {O’Hern},\ and\ \citenamefont
  {Shattuck}}]{isostaticity}%
  \BibitemOpen
  \bibfield  {author} {\bibinfo {author} {\bibfnamefont {S.}~\bibnamefont
  {Papanikolaou}}, \bibinfo {author} {\bibfnamefont {C.~S.}\ \bibnamefont
  {O’Hern}},\ and\ \bibinfo {author} {\bibfnamefont {M.~D.}\ \bibnamefont
  {Shattuck}},\ }\bibfield  {title} {\bibinfo {title} {Isostaticity at
  frictional jamming},\ }\href@noop {} {\bibfield  {journal} {\bibinfo
  {journal} {Phys. Rev. Lett.}\ }\textbf {\bibinfo {volume} {110}},\ \bibinfo
  {pages} {198002} (\bibinfo {year} {2013})}\BibitemShut {NoStop}%
\bibitem [{\citenamefont {Schreck}\ \emph {et~al.}(2012)\citenamefont
  {Schreck}, \citenamefont {Mailman}, \citenamefont {Chakraborty},\ and\
  \citenamefont {O'Hern}}]{Shreck_ellipse}%
  \BibitemOpen
  \bibfield  {author} {\bibinfo {author} {\bibfnamefont {C.~F.}\ \bibnamefont
  {Schreck}}, \bibinfo {author} {\bibfnamefont {M.}~\bibnamefont {Mailman}},
  \bibinfo {author} {\bibfnamefont {B.}~\bibnamefont {Chakraborty}},\ and\
  \bibinfo {author} {\bibfnamefont {C.~S.}\ \bibnamefont {O'Hern}},\ }\bibfield
   {title} {\bibinfo {title} {Constraints and vibrations in static packings of
  ellipsoidal particles},\ }\href@noop {} {\bibfield  {journal} {\bibinfo
  {journal} {Phys. Rev. E}\ }\textbf {\bibinfo {volume} {85}},\ \bibinfo
  {pages} {061305} (\bibinfo {year} {2012})}\BibitemShut {NoStop}%
\bibitem [{\citenamefont {Bitzek}\ \emph {et~al.}(2006)\citenamefont {Bitzek},
  \citenamefont {Koskinen}, \citenamefont {G{\"a}hler}, \citenamefont
  {Moseler},\ and\ \citenamefont {Gumbsch}}]{FIRE}%
  \BibitemOpen
  \bibfield  {author} {\bibinfo {author} {\bibfnamefont {E.}~\bibnamefont
  {Bitzek}}, \bibinfo {author} {\bibfnamefont {P.}~\bibnamefont {Koskinen}},
  \bibinfo {author} {\bibfnamefont {F.}~\bibnamefont {G{\"a}hler}}, \bibinfo
  {author} {\bibfnamefont {M.}~\bibnamefont {Moseler}},\ and\ \bibinfo {author}
  {\bibfnamefont {P.}~\bibnamefont {Gumbsch}},\ }\bibfield  {title} {\bibinfo
  {title} {Structural relaxation made simple},\ }\href@noop {} {\bibfield
  {journal} {\bibinfo  {journal} {Phys. Rev. Lett.}\ }\textbf {\bibinfo
  {volume} {97}},\ \bibinfo {pages} {170201} (\bibinfo {year}
  {2006})}\BibitemShut {NoStop}%
\end{thebibliography}%

\end{document}